\documentstyle[prd,aps]{revtex}

\tighten

\begin{document}

\bibliographystyle{unsrt}

\draft
\title{Local and global gravitational aspects of domain wall space-times}


\renewcommand{\theequation}{\arabic{section}.\arabic{equation}}

\author{ Mirjam Cveti\v c,\footnote{
Electronic address: CVETIC@cvetic.hep.upenn.edu}
Stephen Griffies,\footnote{
Electronic address: GRIFFIES@cvetic.hep.upenn.edu}
\footnote{Address after Aug.\ 1, 1993: Geophysical Fluid
Dynamics Lab, Princeton University, Forrestal Campus, US Route 1,
PO Box 308, Princeton, NJ 08542}
and Harald H. Soleng\footnote{Electronic address:
SOLENG@steinhardt.hep.upenn.edu or
SOLENG@vuoep6.uio.no}
\footnote{On leave from the Institute of Physics, University of Oslo,
P.O. Box 1048 Blindern, N-0316 Oslo, Norway until June 1, 1993}
\footnote{Address after Aug.\ 1, 1993:
NORDITA, Blegdamsvej 17,
DK-Copenhagen 2100 {\O}, Denmark}
}



\address{Department of Physics, University of Pennsylvania,
209 South 33rd Street, Philadelphia, Pennsylvania 19104-6396, USA}

\date{\today}

\maketitle

\begin{abstract}
Local and global gravitational effects
induced by
eternal vacuum domain walls are studied.
We concentrate on thin  walls between non-equal and non-positive
cosmological constants on each side of the wall.
The assumption of
homogeneity, isotropy, and geodesic
completenes of
the
space-time
intrinsic to the
wall as described in the comoving coordinate system
and the constraint that the same symmetries hold
in hypersurfaces parallel to the wall,
yield a general {\em Ansatz\/} for the line-element of
space-time.
We restrict the problem further by demanding
that the wall's
surface energy density,  $\sigma$, is positive
and by requiring that
the infinitely
thin wall represents a thin wall limit of a kink-like
scalar field configuration.
These vacuum domain walls
fall  in three classes depending on the value of their $\sigma$:
(1)\ extreme walls with  $\sigma=\sigma_{{\text{ext}}}$ are planar,
static walls
corresponding  to supersymmetric
configurations, (2)\ non-extreme walls with $\sigma=\sigma_{{\text{non}}}>
\sigma_{{\text{ext}}}$
correspond to expanding bubbles with
observers on either side
of the wall being {\em inside\/} the bubble, and
(3)\
ultra-extreme walls with
$\sigma=\sigma_{{\text{ultra}}}<\sigma_{{\text{ext}}}$
represent
the bubbles of false vacuum decay.
On the sides with less negative
cosmological constant,
the extreme,
non-extreme, and ultra-extreme
walls exhibit
no, repulsive, and attractive effective
``gravitational forces,'' respectively.
These ``gravitational
forces'' are
global effects
not caused by local curvature.
Since the non-extreme wall encloses observers on both sides,
the
supersymmetric system has the lowest gravitational mass
accessable to outside observers. It is conjectured that
similar positive mass protection occurs in all physical systems
and that no finite negative mass object can exist inside the universe.
We also discuss
the global space-time structure of these singularity free
space-times and point out
intriguing analogies with the causal structure of
black holes.
\end{abstract}


\addtocounter{footnote}{-\value{footnote}}


\section{Introduction}

Domain walls are surfaces interpolating between
separate vacua with different
vacuum expectation values of some
scalar field(s).
Such a domain structure can form by the Kibble
\cite{Kibblerev,Vilenkinrev} mechanism whereby different regions of  a hot
universe cool into
different isolated minima of the matter potential.
Domain walls \cite{Vilenkinrev}
can also form as the boundary of a (true)\ vacuum bubble
created by the quantum tunneling process of false
vacuum decay \cite{Coleman}.
Additionally,
the universe could be born through
a quantum tunneling process from nothing
\cite{QuantumCosmology,HartleHawking,Linde}
into different domains with walls in between.

The equivalence between mass and energy in relativistic
physics implies that kinetic energy in
the form of pressure also contributes to the
gravitational mass density. Accordingly, a negative pressure,
that is, a positive tension,
reduces the effective Tolman mass \cite{TOL}
(i.e.\ the gravitational mass)\ of a system.
Domain walls are vacuum-like
hypersurfaces where
the positive tension equals the mass density \cite{Vilenkinrev}, thus
ensuring that  their energy-momentum tensor is
boost invariant in the directions
parallel to the wall.
Taking into account the repulsive
gravitational effect of positive
tension \cite{Gronrep},
a domain wall is by itself a source
of repulsive gravity \cite{Vilenkinrev,IS},  i.e.\ a system with negative
effective gravitational mass.
However, a bubble of
anti-de~Sitter vacuum (AdS$_{4}$)\ which has negative energy
density and positive pressure, is a source of {\em attractive\/} gravity;
it has a positive
effective gravitational mass.
If such a bubble is embedded in a Minkowski (M$_{4}$)\ space-time,
the attractive effect of AdS$_{4}$, caused by its
positive effective gravitational mass,
could
be undercompensated by the domain wall, which has
negative  effective
gravitational mass.
Inertial observers exterior to this object would
find that the wall separating the two vacua accelerates towards them.
Alternatively, it would {\em a priori\/} be possible that the effective
gravitational mass of the  domain wall could be negative enough
to render the total gravitational mass of the
system negative.
Then the result would be a system of total
{\em negative\/} gravitational mass
\cite{Price,Prob,Bonnor}, and
observers
on the side with a less negative cosmological
constant would also be repelled from
the wall.

It has been shown that
{\em ``under very
general assumptions, no static nonsingular solution \ldots exists
for the gravitational field of an uniformly planar matter distribution''\/}
\cite{Dolgov}.  However, if the energy density
of the adjacent vacuum is allowed to be negative,
this result can be avoided. Namely, in this case the positive
effective gravitational
mass density of AdS$_4$ can
precisely cancel the negative effective
gravitational mass density of a domain wall.
Hence, there do exist nonsingular
solutions of Einstein's field equations for
static, planar walls adjacent to
AdS$_{4}$ \cite{Linet,CGRI,CGII}.
Moreover, these solutions are realized as
{\em supersymmetric\/} bosonic field configurations \cite{CGRI}.
In Ref.\ \cite{CGI} a classification of the possible supersymmetric walls
has been given,
and the global
structure
of the space-times induced by these walls has been
explored in
Refs.\ \cite{CDGS,Gibb,Griffies}. Supersymmetric walls
adjacent to Minkowski space are
of particular interest.
The two vacua separated by these walls are {\em degenerate\/} in the
sense that semi-classical
tunneling between them is absolutely suppressed \cite{CGRII}.
The exact cancellation \cite{CDGS} of the gravitational field of the
supersymmetric---hereafter called {\em extreme\/}---domain
wall and the negative vacuum energy of AdS$_{4}$
makes it possible for an observer to be arbitrarily close to such
an infinite planar domain wall
without feeling any gravitational effects.
For this reason, the wall can be viewed as
a perfect
shield for the gravitational field produced by the AdS$_{4}$ vacuum.

As the previous discussion indicates, the effects of
an AdS$_{4}$ region on the
local and global space-time properties of
domain wall systems are significant.
It is of interest to
systematically investigate
the space-times of domain walls
separating regions of non-positive
cosmological constant in Einstein's theory of gravitation.
This study is motivated in part also from the
desire to make a connection of the extreme
walls studied in Refs.\ \cite{CGRI,CGII,CGI,CDGS,Gibb}
in the context of $N=1$ supergravity to
more general non-supersymmetric
configurations.  We also bear in mind that
there are many reasons for
believing the effective theory describing
low energy modes of the superstring is
four dimensional $N=1$ supergravity (see,
for example,  Ref.\ \cite{BruStein}
for further discussion).
The energy densities of
supersymmetric vacua are
strictly non-positive, and thus they induce
AdS$_{4}$ or M$_{4}$
space-times.  Upon breaking supersymmetry, the
vacua can have a
negative, zero, or positive cosmological
constant.

We present a study of the
local and global properties of the space-times
induced by vacuum domain walls between vacua of
arbitrary cosmological constant.
We are primarily concerned with the
domain walls between vacua
of non-positive cosmological constants.
These walls may be classified according to their
energy-density \cite{CGS}.
The non-extreme domain wall
configurations are defined as those with
energy density greater than the corresponding
extreme wall ($\sigma_{{\text{non}}}>\sigma_{{\text{ext}}}$),
and the
ultra-extreme domain walls have
a smaller surface energy density
($\sigma_{{\text{ultra}}}<\sigma_{{\text{ext}}}$) \cite{CGS}.
We start with the {\em Ansatz\/}
that the gravitational field inherits the boost symmetry of the source,
and we demand geodesic completeness of the space-time intrinsic to the
wall as described in the comoving coordinate system,
but we assume nothing about the topology
of the (2+1)-dimensional space-times parallel
to the surface of the domain wall.
It turns out
that non- and ultra extreme walls are non-static
spherical bubbles.
Ultra-extreme solutions
result in an expanding
{\em ultra-extreme bubble\/} which accelerates
towards all time-like outside
observers. This is the tunneling bubble \cite{BKT}
of false vacuum decay \cite{Coleman,CD}.
Solutions with energy density
$\sigma=\sigma_{{\text{non}}}>\sigma_{{\text{ext}}}$
overcompensate the attractive gravity of the inside by
having the energy density larger than the one of the extreme wall;
those are objects with negative effective
Tolman mass.
This does, however,  not yield a negative mass object
inside our universe. Instead,
in the case of vacuum
domain wall bubbles,
gravity warps space-time so that {\em
both\/} sides are on the inside of the non-extreme bubble.
Hence, nature
protects itself against the possibility of objects
with negative total
gravitational mass
by using topology to put all observers on the inside.
This example leads to the conjecture that in analogy with
cosmic censorship preventing naked singularities,
there is also a cosmic positive mass protection.
Note that these non-extreme walls, which are characterized by
energy densities higher than that of the tunneling bubbble,
are examples of
configurations for which supersymmetry provides
a lower bound for the energy-density. Bounds of this type have
also been found in the black hole context
\cite{BlackGib,BlackKal}.
On the other hand, the ultra-extreme solutions with
energy density $\sigma=\sigma_{\text{ultra}}<\sigma_{\text{ext}}$
correspond to objects of
positive effective mass.
Time-like observers on the side with the largest
cosmological constant---observers in the false vacuum---will
inevitably be hit by the inflating bubble.
As seen by inertial observers in the Minkowski space,
the fact that they are hit by the ultra-extreme tunneling bubble
can be understood as a purely kinematic effect with no
connection to gravity whatsoever.  However, according to a Machian school of
thought \cite{Mach} and the {\em general principle of relativity\/}
\cite{GE}, all inertial effects observed in non-inertial frames
may be explained as
due to
the gravitational field of the ``rest of the universe.''
In the present model, using the rest frame of the wall, one
sees freely falling particles outside the bubble being accelerated
inwards, and indeed, from this point of view, the notion
of positive gravitational mass makes sense globally.

The global space-time structure of the
walls considered here
exhibit non-trivial causal structures
due to the presence of AdS$_{4}$ regions
and their associated Cauchy horizons.
Cauchy horizons imply that the
coordinate extensions necessary for providing
geodesically complete space-time manifolds
are non-unique.  Such ambiguities in the geodesic extensions are
reminiscient of those in
the Reissner-Nordstr\"{o}m and
Kerr black hole space-times
\cite{HE,CHAN}.
These space-times form an infinite lattice
\cite{Carter1,Carter2}.
The same lattice
structure is possible in certain
domain wall systems considered here, and
it is even possible to formally associate the
local parameters of the space-times
in the following manner:
the mass, $M$, and
the charge $Q$ (for the Reissner-Nordstr\"{o}m black hole)
or the angular momentum $a$ (for the Kerr black hole)
are associated with the energy density, $\sigma$,  of the
wall and the cosmological constant, $\Lambda \le 0$,
of the adjacent vacuum, respectively.
The non-trivial causal
structure in the domain wall
system is obtained without the space-time
singularities of the black holes.
The absence of singularities opens
the possibility
of studying physics on
causally non-trivial space-times,
with asymptotically Minkowski regions in
certain cases,
without the problems presented
by the black hole curvature singularities.

The paper is organized as follows.  In Sec.\ \ref{grsection},
by working in the comoving frame of the wall,
we deduce the local properties of space-time close to the wall
and the topology of the wall
itself. Then
in Sec.\ \ref{globalsection},
we find geodesic extensions of the comoving coordinate patch and
discuss the global properties of the wall space-times.
It is here that the similarities
in the global structure of the non-extreme domain wall
space-times to the global structure of
black hole geometries are pointed out.
Sec.\ \ref{discussion} contains
a discussion of the results.
Appendix \ref{appencoord} presents
the local coordinate transformations between the
comoving coordinates derived in the text and
the conventional coordinates of the vacuum
AdS$_{4}$, M$_{4}$, and  de~Sitter (dS$_{4}$) space-times.
In Appendix \ref{appenAdS},
we review the salient aspects of
AdS$_{4}$ useful for understanding
the local and global
properties of the walls discussed in this
paper.

\section{local properties of domain wall space-times}
\label{grsection}

\addtocounter{equation}{-\value{equation}}

In this section we present the
local properties of the space-time
induced by a class of vacuum domain walls
in Einstein gravity.
These walls are created from a
scalar field source and separate
vacuum space-times of zero, positive,
and negative cosmological constants.
We shall study explicitly only infinitely thin domain walls, and thus
employ  Israel's
formalism\cite{ISR,ISR2} of singular hypersurfaces.
This method is a familiar
and well defined approach to solving
Einstein's equations in the limit where
the matter source is approximated as an infinitely thin surface.

To begin with, we give general arguments in regard to
the form of the space-time
metric consistent with the symmetries of the
domain wall background.
Then we present Israel's
formalism
to determine the matching conditions
of the space-time at the wall's world tube.
We classify the solutions
and discuss their
effective gravitational fields using Tolman's \cite{TOL}
concept of
gravitational mass.

\subsection{Metric Ansatz}

We solve Einstein's gravitational field equations using
certain symmetry constraints consistent with
properties of a space-time induced by a
domain wall source. In general, the solutions are
time dependent. It is
most convenient to describe the metric in the comoving coordinates  of the
wall system, i.e.\ in the rest frame of the wall.  In this case, the
stress energy associated with the wall is static and depends only on the
coordinate $z$ perpendicular to the wall.

First, we
assume that the spatial part of the metric
intrinsic to the wall
and of the two-dimensional spatial sections
``parallel'' to the wall
are {\em homogeneous\/} and {\em isotropic\/} in the {\em comoving frame\/}
of the wall.
Homogeneity and isotropy reduce the
``parallel'' metric to the spatial part of a (2+1)-dimensional
Friedmann-Lema{\^{\i}}tre-Robertson-Walker (FLRW)
metric~\cite{MTW}.  In the conventional
coordinates this metric has the form
\begin{equation}
(ds_{\parallel})^2=R^2\left[(1-kr^2)^{-1}dr^2+r^2 d\phi^2\right] ,
\label{dstwo}
\end{equation}
where $R$ is independent of the coordinates
$r$ and $\phi$.
The scalar curvature of this surface
is equal to $2k/R^2$.

There are three possible wall geometries.
The first one is a planar wall with $k=0$. In this case
the metric (\ref{dstwo}) can be transformed
to Cartesian coordinates
$(ds_{\parallel})^2 = R^{2} (dx^2+dy^2)$.
The second possibility is a spherical wall with $k>0$. Then the wall is
a closed bubble, in which case
both
$r$ and $\phi$ are
compact coordinates; i.e.,
one may introduce $r=k^{-1/2}\sin\!\theta$
which after a rescaling of $R$ gives the line-element
$(ds_{\parallel})^2=R^2 ( d\theta^2+\sin^2{\!\theta} d\phi^2)$.
Finally, the wall could be a Gauss-B{\'{o}}lyai-Lobachevski
surface with $k<0$. This negatively curved non-compact
surface cannot be embedded in ordinary
3-dimensional Euclidean space,
that is, it cannot be pictured as
an ordinary curved surface \cite{Weinberg}.
Writing $r = (-k)^{-1/2}\sinh\!\varrho$,
with $\varrho>0 $, and rescaling $R$
brings the $k<0$ line element to
$(ds_{\parallel})^{2} = R^2 (d\varrho^{2} +  \sinh^2\!\varrho \,d\phi^{2})$.

For our next assumption,
we demand that the two-dimensional space-time
sections orthogonal to the wall
are {\em static\/}
as observed in the {\em rest frame\/} of the wall.
Hence, if $z$ denotes a coordinate describing the direction
transverse to the wall, and if $t$ represents the {\em proper\/}
time as measured by observers sitting on the wall,
then
$g_{tz}=0$, and both $g_{tt}$ and $g_{zz}$ depend only on $z$.
By an appropriate choice of $z$-coordinate we can write the
orthogonal part of the metric as\footnote{Throughout the paper
we use geometric units of time, i.e. $c\equiv 1$.}
\begin{equation}
(ds_{\perp})^2=A(z)\left(dt^2-dz^2\right) ,
\end{equation}
where $A(z) > 0$.
With $ds^2\equiv (ds_{\perp})^2-(ds_{\parallel})^2$, we get
\begin{equation}
ds^2=A\left(dt^2-dz^2\right)-R^2\left[(1-kr^2)^{-1}dr^2+r^2d\phi^2
\right] ,
\label{Ansatz}
\end{equation}
where $A=A(z)$ and $R=R(t,z)$. The range of $z$ is $z\in\langle -\infty,
\infty\rangle$, and the range of the other coordinates is
like in a FLRW cosmological model \cite{MTW}.

We shall now employ Einstein's equations to
reduce the form of $R(t,z)$.
With the metric (\ref{Ansatz}) and the definition
\begin{equation}
H\equiv {A'\over A},
\label{hdef}
\end{equation}
where $A' = \partial_{z}A(z)$,
the non-trivial components of the
Einstein tensor $G^{\mu}_{\;\;\nu} = {\cal R}^{\mu}_{\:\:\nu}
- {1\over 2} {\cal R}^{\alpha}_{\;\;\alpha} g^{\mu}_{\;\; \nu}$ are
\begin{equation}
  \left.\begin{array}{ccl}
G^{t}_{\;t}&=&
{{k}\over{R^2}}-{{2R''}\over{AR}}-{{R'^2}\over{AR^2}}
+{{HR'}\over{A R}}
+{{\dot{R}^2}\over{AR^2}}  \\
{\mbox{ }}& & {\mbox{ }}\\
G^{z}_{\;t}&=&{{2\dot{R}'}\over{AR}}-{{H\dot{R}}\over{A R}}\\
{\mbox{ }}& & {\mbox{ }}\\
G^{z}_{\;z}&=&
{{k}\over{R^2}}-{{R'^2}\over{AR^2}}
-{{HR'}\over{A R}}+{{2\ddot{R}}\over{AR}}
+{{\dot{R}^2}\over{AR^2}}\\
{\mbox{ }}& & {\mbox{ }}\\
G^{r}_{\;r}&=&G^{\phi}_{\;\;\phi}=
-{{R''}\over{AR}}
+{{\ddot{R}}\over{AR}}
-{{H'}\over{2A}},
         \end{array}
  \right.
\end{equation}
where $\dot{R} = \partial_{t}R(t)$.

Since we are considering matter configurations which
are static in the $(t,z)$-plane, there is no
energy flow in the $z$-direction in the comoving frame of
this coordinate system.
Therefore,
the
$T^{z}_{\;\;t}$
component
of the
energy-momentum tensor vanishes.
Then the $(z,t)$ component of Einstein's field
equations\footnote{The Einstein constant, $\kappa$,
is defined by  $\kappa\equiv 8\pi G$, where $G$ is Newton's
constant.}
\begin{equation}
G^{\mu}_{\;\;\nu} =
\kappa T^{\mu}_{\;\;\nu},
\end{equation}
imply $G^{z}_{\;t}=0$ or
\begin{equation}
2\dot{R}'=H\dot{R} .
\label{zt}
\end{equation}
There are two possible solutions to Eq.\ (\ref{zt}).
In the first case, when the metric is static, $\dot{R}=0$, and
Eq.\ (\ref{zt}) is trivially satisfied.
Let us first focus on the other case when the metric is
non-static.
If $R$ is time-dependent, integration with respect to time yields
\begin{equation}
2R'=HR + M(z)
\label{ztint}
\end{equation}
where $M(z)$ is an arbitrary function of $z$.
A static matter source is defined by a
static stress-energy tensor $T^{\mu}_{\;\;\nu}$, which,
through Einstein's equations, implies a static Einstein
tensor.  A static $G^{r}_{\;r}$ implies
\begin{equation}
{\ddot{R}\over R}={R''\over R}+g(z)
\label{r2dot}
\end{equation}
where $g(z)$ is an arbitrary function of $z$. From the requirement of
time-independence of $G^{t}_{\;t}-G^{z}_{\;z}$ one finds
that
\begin{equation}
\frac{2}{A}\left[-\frac{R''}{R}+\frac{HR'}{R}-\frac{\ddot{R}}{R}\right]
=\frac{2}{A}\left[f(z)-g(z)\right]
\label{fminusg}
\end{equation}
where $f(z)$ is another arbitrary function of $z$.
Using Eq.\ (\ref{r2dot}) and multiplying
Eq.\
(\ref{fminusg})
by $\frac{1}{2}RA$,
we get
\begin{equation}
-2R''+HR'=f(z)R .
\end{equation}
Adding the $z$-derivative of Eq.\ (\ref{ztint}), one finds
\begin{equation}
M'(z)=[H'+f(z)]R.
\end{equation}
This equation holds for any $t$ only if
$M'(z)=0$ and $f(z)=-H'$. Hence, $2R'=HR+M_{0}$ where $M_{0}$ is a constant.

A thick wall solution can be pictured as a stack of infinitely thin
walls. If we require that each of the thin walls are boost invariant along
surfaces of constant $z$,
i.e., each surface of constant $z$ has an
{\em exterior curvature\/} which is boost invariant,
it follows that $M_{0}=0$.
In the static case ($\dot{R}=0$), the
same symmetry
constraint implies $R^2\propto A$.
In general, this symmetry is stronger than just
requiring a boost-invariant $T^{\mu}_{\;\;\nu}$ because it involves
only first order derivatives of the metric coefficents (see
Sec.\ \ref{subsubisrael} below),
whereas $T^{t}_{\;t}=T^{r}_{\;r}$ involves a second order differential
equation.
Also in the thin wall case,
the gravitational field on either side could
have a Kasner type behaviour compensating each other in such a way
that the interpolating singular surface still is boost
invariant \cite{TOM}.
We shall, however,
assume that
the gravitational field
inherits the symmetry of the source, and that
 the directions parallel to the wall are
boost invariant in the strong sense.
The line element
we shall examine is thus given by
\begin{equation}
ds^2= A(z)\left\{ dt^2-dz^2-S^2(t)
\left[(1-kr^2)^{-1}dr^2+r^2 d\phi^2\right]\right\} .
\label{ansatz}
\end{equation}
In summary, the assumptions which imply the form of the metric
 (\ref{ansatz}) are the
following:
\begin{enumerate}
\item
The spatial part of the metric intrinsic to the wall is
{\em homogeneous\/}  and {\em isotropic.\/}
\item
The space-time
section orthogonal to the wall is {\em static.\/}
\item
The directions parallel to the wall
are {\em boost invariant\/} in the strong
sense.\footnote{As shown in Section \ref{subsubvacuum},
boost invariance also puts restrictions
on the functional form of the time-dependence, $S(t)$.}
\end{enumerate}
To solve the Einstein equations for the two
metric functions $A(z)$ and $S(t)$, we follow
the approach of Israel \cite{ISR}
which approximates the wall as infinitely thin in the
$z$-direction.  Thick walls will have the same $S(t)$ as found in
the thin wall approximation and will
asymptotically\footnote{For the familiar kink matter sources,
which we are considering, the solutions typically approach the
thin wall results exponentially fast over its characteristic
length scale.}
approach the
thin wall result for $A(z)$.
For  detailed discussions of this formalism,
with emphasis on domain walls,
see
Refs.\ \cite{IS,SATO,GUTHETAL,BEREZIN}
and Ref.\ \cite{HAR} for an application to
a static plane-symmetric geometry.

\subsection{Thin wall approximation}  \label{subthinwall}

 We assume that the properties of the gravitational field outside a domain wall
can be deduced without knowledge of the internal structure of the wall.
For this reason we employ the thin wall
approximation.
In this approximation, the wall is treated as
infinitely thin, and consequently its energy-momentum tensor
has a $\delta$-function singularity at the wall.
Einstein's field equations imply
that the Einstein tensor also must have a $\delta$-function
singularity here. Because Einstein's tensor is
of second order in derivatives of the metric,
such a singular hypersurface may be modeled by a
metric tensor which has a discontinuity in its first order
derivatives in the
direction transverse to the singular surface.
This idea is the basis of Israel's formalism \cite{ISR}
for singular layers in general relativity.
Furthermore, we shall assume that the wall is a
{\em domain wall,\/} that is, the
wall itself has a vacuum-like surface energy-momentum tensor.
By this we mean a surface energy-momentum tensor which is
proportional to the metric intrinsic to the world tube of the wall.
Physically, one notes that
there is no way to measure velocity relative to a vacuum and thus
a vacuum must have a boost invariant energy-momentum tensor. This leads
to the specific form of the surface energy-momentum tensor.

\subsubsection{Israel's matching conditions} \label{subsubisrael}

Consider a thin wall placed at a {\em constant\/} $z$-coordinate  position in
a space-time described by the metric (\ref{ansatz}).
According to Israel's formalism \cite{ISR},
the surface energy-momentum tensor of the wall is
described by the Lanczos tensor, ${\cal{S}}^{i}_{\;j}$, which is
given by
\begin{equation}
\kappa {\cal{S}}^{i}_{\;j}  = - [K^{i}_{\;j}]^{-}  +
\delta^{i}_{\;j}[K]^{-},
\end{equation}
where $K^{i}_{\;j}$ is
the extrinsic curvature. The  square brackets $[\;\;\;]^{-}$
signify the discontinuity
at the wall placed at $z=z_{0}$, i.e.\
$[\Omega]^{-} \equiv  \Omega_{2} - \Omega_{1}$,
where $\Omega_{2}$ and $\Omega_{1}$ are
the $\epsilon\rightarrow 0$ limit of
$\Omega (z_{0}+\epsilon)$
and $\Omega (z_{0}-\epsilon)$, respectively. The coordinates
$x^{i} \in \{t,r,\phi \}$
describes
the space-time parallel to
the wall.
The extrinsic curvature is given by the covariant derivative of the
space-like unit normal  $n^{\mu}$
of the wall's hyper-space-time:
\begin{equation}
K^{i}_{\;  j}\equiv -n^{i}_{\; ;\, j}\, ,
\end{equation}
where $n^{\mu}$ is specified by the defining relations
\begin{equation}
n^{\mu}n_{\mu}\equiv -1\;\;\;{\rm and}\;\;\;
n^{\mu}u_{\mu}\equiv 0.
\label{ndef}
\end{equation}
Here $u^{\mu}$ is the four velocity of an observer following the
wall. In the present case, $u^{\mu}=\delta^{\mu}_{\;\; t}$
when represented
in the coordinate system of the metric (\ref{ansatz}).
Since the wall is at a {\em constant\/} $z$-coordinate, the
extrinsic curvature takes a simple form if we use a
coordinate system in which the spatial
coordinate transverse to the wall, $\hat{z}$, is normalized so that
$g_{\hat{z}\hat{z}}=-1$,  i.e.\
$d\hat{z}=A^{1/2}dz$. In these coordinates
\begin{equation}
K_{i j} = -{ \zeta \over 2}g_{ij,\hat{z}}\, ,
\end{equation}
where $\zeta = \pm 1$ is a sign factor coming from the
inherent sign ambiguity of the unit normal $n^{\mu}$.
$\zeta$ will be determined
by insisting that the wall
corresponds to the thin limit of a kink-like source.

We choose to scale the coordinates such that $A(z_{0})=1$ where
$z_{0}$ is the position of the wall. Then, without loss of generality
we may perform a global translation of the $z$-coordinate so as to
bring the origin of the $z$-axis to the position of the wall.
In  this way we find
\begin{equation}
\kappa {\cal{S}}^{i}_{\;j}=-\delta^{i}_{\;j}[\zeta H]^{-}_{z=0}
\label{sij}
\end{equation}
corresponding to a vacuum-like equation of state. In other words,
the surface
energy density, $\sigma$, given by
\begin{equation}
\kappa\sigma\equiv \kappa {\cal{S}}^{t}_{\; t}=-[\zeta H]^{-}_{z=0},
\label{sigdef}
\end{equation}
is equal to the wall's tension,
$\tau\equiv {\cal{S}}^{r}_{\;r}={\cal{S}}^{\phi}_{\;\phi}$.

\subsubsection{Vacuum solutions}  \label{subsubvacuum}

To describe the gravitational field exterior to the wall,
we need the solution to Einstein's field equations.
Here we consider thin domain walls interpolating between
two maximally symmetric vacua of zero, positive, or
negative cosmological constant.
Maximally symmetric vacuum solutions to Einstein's theory
are well known \cite{MTW}; nevertheless,
we rederive them below using
the comoving coordinate system of the wall
configurations. The reason to employ
the comoving frame is two-fold:
the first one is technical;
Israel's matching conditions across the wall region
are easily satisfied in this frame.
The second reason is that in this frame the
space-time exhibits cosmological horizons with properties closely
related to the ones of the corresponding  black holes.
In Appendix \ref{appencoord}, we present the local coordinate
relations between the metric in the rest
frame
of the wall and the more conventional
coordinates of maximally symmetric space-times.

Using the metric (\ref{ansatz})
and the definition (\ref{hdef}),
the Einstein-tensor takes the form
\begin{equation}
\left.
\begin{array}{ccl}
G^{t}_{\;t}&=&{1\over A}\left(
{k\over S^2}-H'-{1\over 4}H^2+{\dot{S}^2\over S^2}\right) \\
{\mbox{ }}& & {\mbox{ }}\\
G^{z}_{\;z}&=&{1\over A}\left(
{k\over S^2}-{3\over 4}H^2+{2\ddot{S}\over S}+{\dot{S}^2\over S^2}\right)
\\
{\mbox{ }}& & {\mbox{ }}\\
G^{r}_{\;r}&=&G^{\phi}_{\;\;\phi}={1\over A}\left(
-{1\over 4}H^2-H' +{\ddot{S}\over S}\right).
\end{array}
\right.
\label{Einstein}
\end{equation}
With a static $T^{\mu}_{\;\;\nu}$, and hence static $G^{\mu}_{\;\;\nu}$,
one finds $\ddot{S}/S=q_{0}$
where $q_{0}$ is a real constant.
Then because of boost invariance in the $r$- and $\phi$-directions,
$G^{t}_{\;t}=G^{r}_{\;r}$, and consequently
\begin{equation}
{\ddot{S}\over  S}=q_{0}={\dot{S}^2\over S^2}+{k\over S^2} .
\label{seq}
\end{equation}
With this result, $G^{z}_{\;\;z}=\Lambda$ implies
\begin{equation}
{1\over 4} H^2=q_{0}-{\Lambda\over 3}A ,
\label{aeq}
\end{equation}
where consistency demands $\Lambda A(z) \le 3 q_{0} $.
Note that we shall sometimes parametrize the cosmological constant
by $\Lambda = \pm 3\alpha^{2}$.
Eqs.\ (\ref{seq}) and (\ref{aeq}) are the fundamental equations
yielding the line elements which we study. Note that the $A=1$ solution
of Eq.\ (\ref{aeq})
is a valid solution of the other
components of the Einstein equations only for
the case $q_{0}=\Lambda=0$.

The geometry of hypersurfaces of constant $z$ is determined by solutions
to Eq.\ (\ref{seq}). Without loss of generality,  we normalize the
curvature constant
to $k\in\{-\beta^2,0,\beta^2\}$. The solutions of Eq.\ (\ref{seq})
are classified according to the sign of $q_{0}$.
Thus up to a global translation of the time coordinate,
\begin{eqnarray}
S_{-} &=& \sin \beta t\;\;\;\;\;\;\;\;\;\;   k=-\beta^2
\label{smin}\\
S_{0}\; &=&  \left\{ \begin{array}{ll}
          \beta t \;\;\;\;\;\;\;\;\;  \;       & k=-\beta^2 \\
             1              & k=0
           \end{array}
    \right.
\label{snull}\\
S_{+} &=&  \left\{ \begin{array}{ll}
\sinh \beta t\;\;\,\,  &  k=-\beta^2 \\
 e^{\beta t}     &  k=0        \\
 \cosh \beta t &  k=\beta^2
\end{array}
\right.
\label{splus}
\end{eqnarray}
where the subscripts on $S$ refer to the sign of $q_{0}$.
With these solutions for $S(t)$,
the sections of constant $z$ are (2+1)-dimensional
spaces of maximal symmetry,
i.e., these hyperspaces are
AdS$_{3}$, M$_{3}$ and dS$_{3}$ space-times,
respectively.
Note that the line-element
(\ref{ansatz}) is quadratic in $S(t)$. Hence, the
overall sign of $S$ is arbitrary.  In addition, since
a change in the sign of
$\beta$ is equivalent to time-reversal in all the solutions for
$S(t)$, choosing $\beta\geq 0$ implies no loss of generality.

The solutions for $A(z)$ from Eqs.\ (\ref{hdef}) and
(\ref{aeq}) are the following:
\begin{eqnarray}
A_{-} &= & \begin{array}{ll}
 \beta^2 [ \alpha \cos (\beta z+\vartheta ) ]^{-2}
&
\;\;\;\;\;\;\;\;\;
\;\;\;\;\;\;
\Lambda =-3\alpha^2\leq -3\beta^2
           \end{array}
\label{amin}\;\;\;  \\
A_{0}\;&=&\left\{ \begin{array}{ll}
  ( \alpha z-  1)^{-2}
&
\;\;\;\;\;\;\;\;\;\;
\;\;\;\;\;\;\;\;\;\;
\;\;\;\;\;
\;
\Lambda = -3\alpha^2\\
1     &
\;\;\;\;\;\;\;\;\;\;
\;\;\;\;\;\;\;\;\;\;
\;\;\;\;\;
\;
\Lambda =0
          \end{array}
\right.
\label{anull}\\
A_{+}&=&\left\{ \begin{array}{ll}
\beta^2 [\alpha \sinh(\beta z-\beta z')]^{-2}
 & \;\;\;\;\;\;\,\, \Lambda = -3\alpha^2\\
e^{\pm 2\beta z}
 & \;\;\;\;\;\;\,\, \Lambda =0\\
\beta^2 [\alpha \cosh(\beta z-\beta z'')]^{-2}
 & \;\;\;\;\;\;\,\,  \Lambda =3\alpha^2\leq 3\beta^2
\end{array}
\right.
\label{aplus}
\end{eqnarray}
where the subscripts on $A$ refer to the sign of $q_{0}$.
Without loss of generality we have moved the origin of the $z$-axis
to the position of the wall ($z_{0}=0$).
The three
integration constants
$\vartheta$, $z'$, and $z''$ are
determined by the requirement that $A(0)=1$. This
normalization yields
\begin{eqnarray}
\vartheta_{\pm}&=&\pm {\mbox{arccos}} (\beta/\alpha)\\
\beta z'_{\pm} &=&
\frac{1}{2}\ln \left[1+ {2\beta^2\over\alpha^2}\pm {2\beta\over\alpha^2}
(\alpha^2+\beta^2)^{1/2}\right] \label{deltadet}\\
\beta z''_{\pm} &=&
\frac{1}{2}\ln\left[-1+ {2\beta^2\over\alpha^2}\mp {2\beta\over\alpha^2}
(\beta^2-\alpha^2)^{1/2}\right]. \label{gammadet}
\end{eqnarray}
The constants $\beta z'$ and $\beta z''$ satisfy
$e^{2\beta (z'_{+}+z'_{-})}=
e^{2\beta (z''_{+}+z''_{-})}=1$
and $e^{2\beta z''_{-}}>1 > e^{2\beta z''_{+}}$ and
$e^{2\beta z'_{-}}\le 1 \le e^{2\beta z'_{+}}$
where
in the last case,
equality is obtained
when $\beta = 0$. As one can see from Eq.\ (\ref{gammadet}),
there is
no extreme limit ($\beta\rightarrow 0$)\ in the de~Sitter case.

Recall that the solutions (\ref{amin})--(\ref{aplus}) are the form for
the metric function $A(z)$ some distance away from a wall
centered at $z=0$.
In the $A_{-}$ solution (\ref{amin}),
$z$ is an angular coordinate,
but as seen from Eq.\ (\ref{smin}), the metric has singularities for certain
values of comoving time, $t$, and because of this, this solution will
not be further discussed here.

Points where $A(z)$ is singular represent affine boundaries
of the space-time.  They are an infinite proper distance away
from every other point within the particular
interval of $z$ considered.
The first of the $A_{0}$ solutions (\ref{anull})
is the line element for AdS$_{4}$
written in the so-called {\em horo-spherical\/} coordinate system.
Discussions of these coordinates can be found in
Refs.\ \cite{CGI,Gibb,Griffies}.
For the first two $q_{0}=\beta^{2}$ solutions,
one is able to consider both singular or non-singular
functions $A(z)$ depending on the choice of $\pm \beta$ or $\beta z'_{\pm}$
for the respective solutions $\Lambda = 0$ or $\Lambda= -3\alpha^{2}$.
For example, if we choose $A=e^{ +2\beta z }$ for
$z>0$ (the M$_{4}$ side of a wall), there will be a coordinate singularity
at $z=\infty$ corresponding to the null boundary of the space-time.
Choosing the decaying exponential will cause $z=\infty$ to be a
finite proper distance away from any other point;  thus
the coordinate patch must be extended.
Likewise, choosing $\beta z'_{-}$ for $z<0$ (the AdS$_{4}$ side)\
will result in a time-like coordinate singularity at $z=z^{*}$, where
$z^{*}=z'_{-}$.
This coordinate singularity is
again the boundary of the space-time,  only now it is time-like
rather than null as in the M$_{4}$ case.
The de~Sitter solution  is always  without coordinate singularities.
Geodesic extensions
of the domain wall space-times will be further discussed
in Sec.\ \ref{globalsection}.

\subsubsection{Topology of the walls}

On account of boost invariance in the directions parallel to the wall,
the spatial curvature of constant $z$ sections is not
unambiguously defined. Since the wall is
homogeneous and boost invariant,
there is no preferred frame in the (2+1)-dimensional
space-time of the wall. Observers can measure the
curvature of space by
sending light signals to each other and
measuring angles of triangles,
but the results will depend on the
relative motion of these observers.
For instance, the two $S_{0}$ solutions (\ref{snull})---the
Milne type solution with
$S=\beta t$ and $k=-\beta^2$ and the inertial
Minkowski solution with $S=1$ and $k=0$---are related by a
coordinate transformation \cite{Robertson} not
involving the transverse coordinate $z$ and
therefore describe locally
equivalent space-times.\footnote{The same holds in
(3+1)-dimensions where a special set of observers see
flat Minkowski space-time as the expanding Milne
universe
\cite{Robertson} with hyperbolic spatial sections.}
Similarly, the constant $z$ sections of
the three $S_{+}$ solutions (\ref{splus}) all represent
(2+1)-dimensional de~Sitter space-time (dS$_{3}$).
The topology of dS$_{3}$ is
${\bf R}({\text{time}})\times {\bf S}^{2}({\text{space}})$.
Embedded in a flat higher-dimensional Minkowski space-time,
dS$_{3}$ represents a hyperboloid \cite{Gibb},
and the three possible spatial curvatures correspond
to three different choices of constant time slices of this
hyperboloid \cite{HE,Robertson}.
Just as in the four-dimensional case \cite{HE,Robertson},
only the positive curvature solution gives a complete
covering of dS$_{3}$.

Therefore,  of the three $S_{+}$ solutions in Eq.\ (\ref{splus}),
only the one corresponding to a
compact {\em spherical wall\/} with radius  $\beta^{-1} A(z)^{-1/2} S(t)$
completely covers
the constant $z$ space-time.
The $S_{0}$ solution is the {\em only\/} wall which
represents a noncompact {\em planar\/} ($k=0$) wall.
In summary,  homogeneity, isotropy, and  geodesic completeness of the
space-time intrinsic to the wall as described in comoving
coordinates impose
constraints on the topology of the domain wall;
it either corresponds to a {\em
static, planar\/} wall or to a {\em time dependent bubble\/}:
\begin{equation}
   S_{0}=1 \;{\text{ and }}\;k=0\;\;
  {\text{ {\em or} }}\;\; S_{+}=\cosh\beta t\;{\text{ and }}\; k>0.
\label{relevantsolutions}
\end{equation}

Note again that the solutions  $S(t)$ are valid in the thick wall
case because its functional form
is determined by the requirement of boost invariance alone
and thus independent of $A(z)$.
The solutions for $A(z)$ in
Eqs.\ (\ref{anull})--(\ref{aplus}) approximate
the local form of the metric some distance away from
the wall.  To describe the space-time on both sides of
the wall centered at  $z=0$, we fix the
same value for the parameter $q_{0}$ and  the curvature $k$
on both sides and choose any of the corresponding solutions
for $A(z)$.  Consequently, there are two
distinct wall configurations
described by the line element (\ref{ansatz}):
one whose spatial topology parallel to the wall
surface is ${\bf R}^2$ and
another whose topology is ${\bf S}^2$.

We would like to elaborate on the
closed bubble solution ($k=\beta^{2}$, $S(t) = \cosh\beta t$).
First, we transform to canonical ${\bf S}^{2}$ coordinates
in Eq.\ (\ref{ansatz}).
Writing $\beta r=\sin\!\theta$, brings
the metric
to
\begin{equation}
ds^{2} = A(z)\left[ dt^{2} - dz^{2} -
\beta^{-2}\cosh^{2}\!\beta t\;
( d\theta^{2} + \sin^{2}\theta d\phi^{2})\right] .
\end{equation}
For these solutions, {\em both\/}  sides of the wall
have a compact spherical line element in the spatial directions
parallel to the wall.
Thus, observers on {\em each\/} side of the wall
observe the bubble wall
collapsing from an infinite radius at $t=-\infty$ to a
finite radius at $t=0$
and then re-expanding to an infinite radius at $t=\infty$.

This may sound strange since the wall is located at a fixed
$z$-coordinate. However, the proper surface area
of the bubble is given by
$4\pi R_{b}^2 \equiv 4 \pi \beta^{-2} A(z) S^{2}(t)$,
which means that the radius as
measured by this formula is time-dependent.
Geometrically, $R_{b}$ is the {\em radius of curvature\/} of the wall;
physically, it is a measure of the proper size of the bubble.
Hence, the bounce description
is not a coordinate artifact.
Indeed, observers sitting on
the bubble and measuring its
surface area with {\em standard\/} measuring
rods extending around the bubble,
versus time as measured by {\em standard\/}
clocks, will observe the bounce directly.

Notice also that for the bubble solution,
the direction orthogonal
to the wall is described by the $z$-coordinate, which has
the range $\langle-\infty,\infty\rangle$.
This is twice the range of a radial
coordinate in Euclidean space. However,
it is $R_{b}$ rather than $z$ which plays the r{\^o}le of the radius.
Moreover, when looking at the $z$-dependence of $R_{b}$ at constant $t$,
$R_{b}$ can either  decrease on both sides of the bubble or
increase on one side and decrease on the other.
In the former case,
using
measurements of
angular distance,
observers on both sides of the wall
can
rightfully
say that they are on the {\em inside\/}  of the bubble.
In this picture
$z\in \langle-\infty,0]$
and $z\in [0,\infty\rangle$
map onto twice $R_{b} \in [0,R_{0}]$,
where $R_{0}= \beta^{-1} S(t)A^{1/2}(0)$,
i.e., the infinite range of
$z$ maps onto two
ranges for the angular distance corresponding
to two spheres (the two insides of the bubble).
In the latter case,
$z\in \langle-\infty,0]$
and $z\in [0,\infty\rangle$
map onto $R_{b} \in [0,R_{0}]$ and $R_{b} \in [R_{0},\infty\rangle$,
corresponding to the {\em inside\/} and the
{\em outside\/} of a single bubble.
As we shall see later on, there is no physical bubble solution
(no solution with surface energy
$\sigma > 0$)
corresponding to a space where both observers are on the outside of
the bubble.
We conclude that a spherical vacuum domain wall
has at least {\em one\/}
inside.

For the $q_{0} = 0$ and $q_{0}=\beta^{2}$
wall solutions,  the constant $z$
sections correspond to M$_{3}$ and dS$_{3}$, respectively.
The novel causal structure for the walls,
which we shall discuss in Sec.\ \ref{globalsection},
involves only the (1+1)-dimensional
sections orthogonal to the walls.
This structure is deduced from an analysis of the $(t,z)$
components of the line elements.

\subsubsection{Surface energy density of the domain walls}
\label{subsubenergydensity}
By matching vacuum solutions with the
same $k$ and $q_{0}$ and using
Eqs.\ (\ref{sigdef}) and (\ref{aeq}), we find that the
interpolating domain walls have the following
surface energy density, $\sigma$, and tension, $\tau=\sigma$:
\begin{equation}
\kappa\sigma = 2\zeta_{1}h_{1}\left(q_{0}-{\Lambda_{1} \over 3}\right)^{1/2}
-2\zeta_{2}h_{2}\left(q_{0}-{\Lambda_{2}\over 3}\right)^{1/2} ,
\label{gensig}
\end{equation}
where the indices $1$ and $2$ stand for values at $z<0$ and $z>0$,
respectively.

In the thin wall formalism, there is an ambiguity in the sign of the
unit normal $n^{\mu}$ defined in Eq.\ (\ref{ndef}).
Also in going from Eq.\
(\ref{aeq}) to Eq.\ (\ref{gensig}), we pick up
another sign ambiguity because Eq.\ (\ref{aeq}) is quadratic in $H$.
These sign ambiguities are taken care of by the sign factors
$h_{i}=\pm 1$ and
$\zeta_{i}=\pm 1$ with $i\in\{1,2\}$.
The first sign factor, $h_{i}$, is determined as follows.
If $A_{i}$ is an increasing function
of $z$, then $h_{i}=1$, and conversely if $A_{i}$ is decreasing then
$h_{i}=-1$.
Physically relevant solutions to the matching conditions involve
those with a {\em positive\/}  energy density, $\sigma$, as well as
sources corresponding to the infinitely thin wall limit of a
kink-like source.
Associating the direction of the wall's outward normal with a
chosen direction of the matter source gradient
implies $\zeta_{1} =  \zeta_{2}=1$ for a kink-type source.
Sources with $\zeta_{1} = -\zeta_{2}$
correspond to spike-type sources which we do not consider.
We will chose to orient the $z$-coordinate so that
the
vacuum of lowest energy (most negative $\Lambda$)\ will be placed
on the $z<0$ side.

\subsection{Classification of the domain wall solutions}
\label{subclassification}

The previous solutions can be classified according to the
three values of the parameter $q_{0}$.
The metrics written here in the comoving frame
are locally related to more standard
coordinates of M$_{4}$, AdS$_{4}$, and dS$_{4}$,
depending on the cosmological constant $\Lambda = 0,
\Lambda < 0,$ or $\Lambda > 0$, respectively.  We
exhibit these relations in  Appendix \ref{appencoord}.

In the case $q_{0} = -\beta^{2}$, the
line element intrinsic to the wall has periodic singularities
in comoving time.
Therefore, we do not consider it further here.
The possibilities $q_{0} = 0$ and $\beta^{2}$
are classified in the following two subsections.

\subsubsection{Extreme walls $(q_{0}=0)$ }

The $q_{0}=0$ solutions
exist for AdS$_{4}$ $(\Lambda = -3\alpha^{2})$ and for
M$_{4}$ $(\Lambda = 0)$, the latter being the
$\Lambda\rightarrow 0$ limit of the former.
If the vacua between which the domain wall interpolates
are supersymmetric, then these walls are
realized as supersymmetric bosonic configurations in
$N=1$ supergravity coupled to chiral matter superfields~\cite{CGRI}.
In this case, the walls are called {\em extreme\/} domain walls~\cite{CDGS}
in analogy with the extreme Reissner-Nordstr{\"{o}}m black hole,
which is also realized as a supersymmetric bosonic
configuration~\cite{GibHull}.

The line element (\ref{ansatz})  for the $q_{0} = 0$ case
has two physically distinct solutions:
$S=1$ and the two solutions for $A(z)$ of Eq.\ (\ref{anull}).
The solutions are written in canonical
M$_{4}$ Cartesian coordinates and {\em horo-spherical\/} AdS$_{4}$
coordinates, respectively.
The horo-spherical coordinates
are natural for describing AdS$_{4}$ when it
is juxtaposed with a flat M$_{4}$ region, as
it is in the $k=0$ extreme wall \cite{CDGS,Gibb,Griffies}.

The extreme walls \cite{CGS}
have been classified into three types according to
the nature of their space-times \cite{CGI}.
Type I is the planar wall
interpolating between M$_{4}$ and AdS$_{4}$. Type II walls
interpolate between two
AdS$_{4}$ regions with the metric conformal factor $A(z)$
becoming $(\alpha_{i} z)^{-2}$ on the respective sides ($i\in\{1,2\}$)
of the wall. Type III walls
interpolate between two AdS$_{4}$ spaces with different
cosmological constants in such a way that the conformal
factor increases without bound
for $z$ moving away from the wall on one side.
The singularity in $A(z)$ represents the time-like
boundary of the space-time,  i.e.\  affine infinity.
In the underlying supergravity theory, Type I, II, and III walls
can also be distinguished by the behaviour of the
superpotential as the wall interpolates
between the two supersymmetric vacua~\cite{CGI}.

The energy density of the extreme walls is
\begin{equation}
\kappa\sigma_{\text{ext}}=2(\alpha_{1}\pm\alpha_{2}),
\label{ssex}
\end{equation}
where $\Lambda_{i}=-3\alpha^{2}_{i}$ with the plus sign for Type II walls
and the minus sign for Type III walls. It is understood that
$\alpha_{1}>\alpha_{2}$ in the latter case. Type I corresponds to
$\alpha_{2}=0$.
By using the thin wall approximation,
a reflection symmetric
special case of the Type II walls was
independently found by Linet~\cite{Linet}.

\subsubsection{Non- and ultra-extreme walls $(q_{0}=\beta^2)$}

The $q_{0}=\beta^2$ solutions
exist for closed walls for all values of the
cosmological constant.
Specifically, we have the four
physically distinct
solutions (\ref{aplus}) for $A(z)$.

A non-extreme wall has an energy density higher than
the corresponding extreme wall.
Explicitly, the energy density
is
\begin{equation}
\kappa\sigma_{\text{non}}=2\left(\pm\alpha^2_{1}+\beta^2\right)^{1/2}+
2\left(\pm\alpha^2_{2}+\beta^2\right)^{1/2}.
\label{nex}
\end{equation}
Here $\Lambda_{i}=\mp3\alpha_{i}^2$, so that a minus  sign in front
of $\alpha^2$  in the above expression corresponds to
a {\em positive\/} $\Lambda$-term.
In the de~Sitter case
$\alpha^{2}_{i} \leq \beta^{2}$ \cite{SATO}.

The ``planar'' reflection symmetric wall
discussed by Vilenkin \cite{Vil} and Ipser and Sikivie \cite{IS}
with $\Lambda_{1}=\Lambda_{2}=0$
is a special non-extreme wall.
These walls have
an energy density $\kappa\sigma=4\beta$.
Thick wall generalizations
have been studied in Refs.\ \cite{Goetz,Mukharjee}.
Note that these walls are
spherically symmetric bubbles \cite{Gibb}
rather than planar walls.
For all the non-extreme walls, the
radius of curvature $R_{b}$ of concentric shells at
constant time {\em decreases\/}
away from the bubble on {\em both\/} sides. Hence,
in this sense, the
non-extreme bubbles have {\em two\/} insides.

An ultra-extreme wall has an energy density lower than
the corresponding extreme wall. Its
energy density
is
\begin{equation}
\kappa\sigma_{\text{ultra}}=2\left(\pm\alpha^2_{1}+\beta^2\right)^{1/2}-
2\left(\pm\alpha^2_{2}+\beta^2\right)^{1/2}.
\label{uex}
\end{equation}
The signs in front of $\alpha^2$ correspond to
$\Lambda_{i}=\mp 3\alpha^2_{i}$. It follows that
if $\Lambda_{i} = 3\alpha^{2}_{i}$,
then $\alpha^{2}_{i} \leq \beta^{2}$ \cite{SATO}.

Ultra-extreme AdS$_{4}$--AdS$_{4}$ and AdS$_{4}$--M$_{4}$ bubbles of the
false vacuum decay  are more like
ordinary
bubbles than non-extreme bubbles;
their radii increase away from the wall on one side
and decrease on the other. Thus, they
have one inside and one outside, the outside being the higher
energy vacuum.
M$_{4}$--dS$_{4}$ and dS$_{4}$--dS$_{4}$  walls were addressed in
Refs.\cite{SATO,GUTHETAL,BEREZIN}.

In the static or extreme limit $\beta\rightarrow 0$,
the non-extreme and ultra-extreme walls of AdS$_{4}$--AdS$_{4}$ walls
reduce to the
Type II and Type III extreme walls, respectively.
The non- and ultra-extreme bubbles were discussed
by Berezin, Kuzmin and Tkachev
\cite{BKT} with particular emphasis on the ultra-extreme
bubbles corresponding to false vacuum decay.

\subsection{Fiducial observers and Tolman's mass for extreme walls
of Type I and Type II}
\subsubsection{Fiducial observers}

Inertial observers on the
M$_{4}$ side of the extreme AdS$_{4}$--M$_{4}$ wall
experiences no gravitational effects\footnote{Strictly speaking the
``gravitational force'' is exponentially
close to zero (exactly zero in the thin wall approximation),
which is consistent with the appellation ``Minkowski side.''}
from the infinite wall.
A particular way to understand this result is to investigate the
proper acceleration,\footnote{Hats denote tensor components
relative to a (pseudo)-orthonormal tetrad frame, i.e.\ a
physical frame.}
$a^{\hat{\mu}}$,
necessary to be a fiducial observer (an observer at a fixed spatial
position)\ in the space-time described by the conformally flat
metric $A(z)(dt^{2} - dx^{2} - dy^{2} - dz^{2})$.
This acceleration
is given by \cite{CGI}
\begin{equation}
   a^{\hat{\mu}}=\frac{1}{2} A^{-1/2} H \delta^{\hat{\mu}}_{\;\; \hat{z} }
\end{equation}
and is directed {\em towards\/}
the wall region thus exhibiting
``repulsive gravity.''
Clearly for $A(z) = 1$ the acceleration is
zero, and no gravitational effects are felt.

On the AdS$_{4}$ side, $A(z) = (\alpha z -  1)^{-2}$,
which yields a constant proper acceleration of magnitude $\alpha$
for
fiducial observers. Observe that this acceleration is {\em half\/}
the surface mass
density of an AdS$_{4}$--M$_{4}$ wall.
This fact will prove
important in the following formulation of the Tolman mass per
area for the domain wall system.

\subsubsection{Tolman's mass per area}

A useful way to understand the equilibrium between the
extremal wall and the adjacent space-times is to compute Tolman's effective
gravitational mass for the system. In spaces of high symmetry
it is possible
to rewrite Einstein's field equations as an integral which can be
interpreted as an expression for an effective gravitational mass.
By computing this mass one can understand the effective
``gravitational forces'' and their sources.

For example, in pure AdS$_{4}$, the effective gravitational mass per
volume
is positive, thus indicating the attractive nature of the gravitational
field produced by the negative energy vacuum.
As discussed in
Appendix \ref{appenAdS}, the motion of test particles is oscillatory
in pure AdS$_{4}$ reflecting the fact that every point in the
space-time attracts the particles with its
positive effective mass density.  The converse holds for pure dS$_{4}$,
which is a familiar result from inflationary
cosmology
where the repulsive nature of a positive energy vacuum drives the universe into
exponential expansion.

In the domain wall system, the relevant
object is the gravitational mass per area.
In particular, the zero Tolman's mass of the extreme
AdS$_{4}$--M$_{4}$ domain wall
enables one to understand why it is
possible to be on the M$_{4}$ side and near the
infinite wall,
which possesses a non-zero mass
density, and yet feel no gravitational
effects.

Tolman's formula for the gravitational mass
was originally derived
for a static spherically symmetric metric \cite{TOL}.
We generalize this result to
the case of the static planar symmetry of the Type I and II extremal walls.
In the derivation of Tolman's mass formula one focuses on the
{\em generalized surface gravity,\/}\footnote{In black hole
theory \cite{Thorne}
the term
{\em ``surface gravity''\/}
means the acceleration at
the horizon surface. By the {\em ``generalized surface gravity''\/}
we mean the acceleration of gravity at any surface.}
that is, the gravitational
acceleration
as measured with {\em standard\/}
rods and {\em coordinate\/} clocks \cite{Gron}
\begin{equation}
k^{i} \equiv -\sqrt{g_{tt}}a^{\hat{\imath}} .
\end{equation}
Since we are seeking a
``Newtonian'' concept of gravitational mass,
it is natural to use {\em coordinate\/} time instead of the local proper
time because the latter time is
subject to time-dilation effects,
which we eliminate by the factor of $\sqrt{g_{tt}}$.
Note that only in the case where this coordinate time becomes the
proper time of the observers (e.g.\ at infinity)\
is one guaranteed
a direct physical interpretation of
this mass. In the Schwarzschild case,
for instance, the Tolman mass obtained by integrating from the origin to the
surface of a finite, static source,
is identical to the $M$-parameter in the Schwarzschild metric
describing the vacuum exterior to the source.
For a metric as given in Eq.\ (\ref{ansatz}), with $S(t) = 1$,
the
generalized surface gravity is
\begin{equation}
   k^{i}=-\frac{1}{2}H\delta^{i}_{\; z}\, .
\label{kres}
\end{equation}
In a spherically symmetric
four-dimensional space-time, one relates $k^{r}$ to
the Tolman mass by defining  a mass $M$ in such a way that the Newtonian
force law is reproduced: $GM\equiv -r^2 k^{r}$, and in the
three dimensional case $G_{3}M_{3}\equiv -rk^{r}$, where
$G_{3}$ is the (2+1)-dimensional Newton's constant \cite{Soleng}.
In the plane-symmetric four-dimensional case, considered here,
 we deal with
an essentially two-dimensional problem, and   the appropriate Newtonian
force law  implies $\kappa\Sigma\equiv -2k^{z}$ where $\Sigma$ is the
gravitational mass per area of the plane.
The factor of two is included because the gravitational acceleration is
half the mass per area
in the planar symmetric case (in the reflection symmetric case \cite{Vil,AG}
one finds that the acceleration on both sides is a quarter of the mass
density).
In the same spirit as the compact
spherically symmetric case, Eq.\ (\ref{kres}) leads to
\begin{equation}
\kappa\Sigma (z)= H(z) .
\label{massden}
\end{equation}

For this equation to make sense in the wall case,
we rewrite the right hand side
in terms of an integral.
Starting from the Einstein tensor (\ref{Einstein}),
one finds  (in the static case $q_{0}=0$)
\begin{equation}
G^{t}_{\;t}-G^{z}_{\;z}-G^{r}_{\;r}-G^{\phi}_{\;\;\phi}=\frac{A''}{A^2} .
\label{tolmanI}
\end{equation}
Using
$\sqrt{-g^{(4)}}=A^2$, $\sqrt{g^{(2)}}=A$,
$G^{\mu}_{\;\;\nu} = \kappa T^{\mu}_{\;\;\nu}$, and
Eq.\ (\ref{tolmanI}),
we can rewrite the righthand side of Eq.\ (\ref{massden}) as the following
integral expression
\begin{equation}
   \kappa\Sigma (z) = H(z) =  \frac{ \kappa\int_{-\infty}^{z}
   \sqrt{-g^{(4)}}dz'
   \left(
   T^{t}_{\;t}-T^{z}_{\;z}-T^{r}_{\;r}-T^{\theta}_{\;\;\theta}
   \right)
   \int dx dy }
   { \sqrt{g^{(2)}(z)} \int dx dy }
\label{Tolman}
\end{equation}
where we used
$A'(-\infty)=0$,  as is the case for both the
asymptotically M$_{4}$ and AdS$_{4}$ sides of the Type I and II
extremal walls.
The numerator on the right hand side of
Eq.\ (\ref{Tolman})
is recognized as the Tolman mass of a static space-time\cite{TOL}.
This mass is non-local, which is consistant with it giving
a Newtonian perspective to a static space-time.
Basically, the Tolman mass formula expresses the fact that
mass and energy are equivalent quantities in relativistic physics.
On account of this,
energy in the form of pressure contributes to the gravitational field
along with the mass density. In this vein, one can define
a gravitational mass density by $\rho_{g}\equiv\rho+3p$
for a perfect fluid in $3+1$ dimensions.
In the limit where we integrate from $z=-\infty$ to $z=\infty$, we get
$\Sigma(\infty)=0$. This means that
the total gravitational mass of
the Type I and II
extreme space-times is zero.
It should be noted that the Type III space-time is
causally identical to pure AdS$_{4}$
(see Sec.\ \ref{globalsection}).
Therefore, the effective mass per volume of this
system is the relevant object; the effective mass
per area, as with pure AdS$_{4}$, is infinite.

\subsubsection{Tolman's mass per area of a thin extreme wall}

In the thin wall approximation, we can distinguish contributions to
the Tolman mass per area
due to the wall itself and due to the vacuum energy of
the adjacent
space-time.
In this case, a domain wall
has an effective  gravitational  mass per area, $\Sigma_{{\rm wall}} =
S^{t}_{\;\; t} - S^{r}_{\;\;r } - S^{\phi}_{\;\; \phi}$, given by
$\Sigma_{{\rm wall}} \equiv \sigma - 2\tau$.
Since the tension, $\tau$, is equal to the energy density, $\sigma$,
for a vacuum domain wall
(see Eqs.\ (\ref{sij}) and (\ref{sigdef})),
we find $\Sigma_{{\rm wall}} = -\sigma<0$.
By use of Eq.\ (\ref{ssex}), one finds that $\kappa\sigma=2\alpha$, which
yields
\begin{equation}
\kappa\Sigma_{{\rm wall}} = - 2 \alpha .
\label{gravitymass}
\end{equation}
This negative gravitational mass per area for the wall,
with its {\em repulsive\/}
gravity,
must be compensated by a positive gravitational
surface mass density  from the AdS$_{4}$ space-time on the AdS$_{4}$
side of the wall in order for there to
be no force on the M$_{4}$ side.
This is precisely the case as we now show.
Again, taking into account the effect of vacuum pressure, $p_{v}=-\rho_{v}$,
the
gravitational mass density of AdS$_{4}$  is
\begin{equation}
\kappa\rho_{g} =  \Lambda - 3\Lambda = 6\alpha^{2} .
\label{adsenergi}
\end{equation}
Integrating out the $z$-direction from
$z= -\infty$ to the
position of the wall at
$z=0$
yields the
following mass per area for the AdS$_{4}$ side of the wall:
\begin{equation}
\kappa\Sigma_{AdS} = \lim_{ z  \rightarrow 0 }
\left[ {  \int_{-\infty}^{z} (6 \alpha^{2})
\sqrt{-g^{(4)}}  dz\int dx dy \over \int \sqrt{g^{(2)}}dx dy  } \right]=
\lim_{z\rightarrow 0}\left[-{2\alpha\over (\alpha z-1)}
\right]=
2\alpha  .
\label{surfaceenergy}
\end{equation}
Hence, as seen from the M$_{4}$ side of the domain wall
(the $z > 0$ side), there
are two gravitational surface mass densities on the
$z \le  0$ side. Firstly, there is a
{\em negative\/} mass per area coming
from the domain wall: $\kappa\Sigma_{\rm{wall}} = -2\alpha$.
Secondly, there is a {\em positive\/} integrated  mass per area coming from
AdS$_{4}$  space itself: $\kappa\Sigma_{AdS} = 2\alpha$,
which exactly cancels
that of the domain wall.

The analysis used for the extreme AdS$_{4}$--M$_{4}$ wall can
also be applied to  the extreme type II AdS$_{4}$--AdS$_{4}$ wall.
When one side is M$_{4}$,
the Killing time, $t$, corresponds to the
proper time of an observer infinitely far away from the wall
on the Minkowski side. In the Type II case, one may use an observer sitting
in the center of the wall. Here too, there
is a frame where
all the connection coefficients vanish, the metric is Minkowskian,
and where the proper time of the observer is equal to the Killing time.
Thus, in the thin wall approximation,
one
finds the effective
mass per area of the two AdS$_{4}$ sides to be $2(\alpha_{1} + \alpha_{2})$.
This positive effective mass is exactly cancelled by the negative
effective mass of the domain wall separating the two regions of AdS$_{4}$.
Likewise, the general expression Eq.\ (\ref{Tolman}) yields a
zero Tolman mass per area for the space-time.

Note that in the above calculations we have integrated along a
constant time slice $-\infty < z < \infty.$
As we shall see when discussing the global space-time induced
from the extreme domain walls in Sec.\ \ref{globalsection}, there is a past and
future Cauchy horizon
for data placed on such a slice.
The above calculation implicitly assumes no
contribution to the effective mass arising from the
past of the past Cauchy horizon.  This assumption is
consistent with the extensions of the space-time
beyond the Cauchy horizon considered in Sec.\ \ref{globalsection}.
Indeed, it is the only assumption consistent with there
being a global balance of gravitational ``forces.''

\section{global properties of domain wall space-times}
\label{globalsection}

The line elements found in the previous section are solutions
to Einstein's equations under the chosen assumptions, but
the global structure of the space-times they describe is not
prescribed since the field equations are local.  In this section,
we present geodesically complete space-times induced from the
domain walls.  In particular, we give the
conformal diagrams
and the corresponding coordinate atlases.  As we shall see,
the resulting geodesically complete space-times exhibit
non-trivial causal structure. This structure is
achieved without space-time singularities.
The most symmetric extensions possess a
lattice structure similar to those
of the extreme and non-extreme
Reissner-Nordstr\" om and Kerr black holes.

We begin this section with a discussion of the global
space-time for the three extreme walls.
Then, we present extensions for the non- and ultra-extreme
 wall
space-times.
Global space-times for the non- and ultra-extreme walls were
described in Ref.\ \cite{CGS} for the case with
M$_{4}$ on one side and AdS$_{4}$ on the other and in
Ref.\ \cite{Gibb} for the M$_{4}$--M$_{4}$ case.
Here we also present the walls with AdS$_{4}$ on
both sides; i.e.\ the non-  and ultra-extreme
generalizations of
the Type II and Type III
walls.\footnote{We do not discuss the global space-times of the
walls with dS$_{4}$ on at least one side.  Such walls can arise
from quantum tunneling when $\Lambda_{1} \ne \Lambda_{2}$
and have been discussed in Refs.\ \cite{BKT,SATO,GUTHETAL,BEREZIN}.
The fine-tuned $\Lambda_{1} = \Lambda_{2} > 0$ will not be discussed
here either.}

As indicated earlier,
the parameter $\beta$  in the parametrization of the
metric represents the
deviation of the field configuration from the corresponding
supersymmetric (extreme)\ one.
In addition,
the spatial part of the metric
internal to the
wall
($k=\beta^{2}$) is geodesically complete.
The curvature
scalar of this part of the
space-time is $2\beta^{2}A(z)^{-1}S(t)^{-2}$.
Therefore, only in the extreme $\beta \equiv 0$ case do we have
a non-compact {\em planar\/} configuration.  For $\beta \ne 0$,
the walls are {\em compact\/} bubbles \cite{Gibb}.
The (2+1)-dimensional space-time
parallel to the wall is the (2+1)-dimensional
de~Sitter space (dS$_{3}$).  The causal structure of
dS$_{3}$ is similar to the well known
dS$_{4}$, which is discussed in Ref.\
\cite{HE}. In the following
we therefore focus on the novel  aspects of the
direction transverse to the wall,
that is, we consider geodesic extensions
in the $(t,z)$ directions.

Understanding of the space-times
induced by these configurations
is facilitated by examining the causal structure of pure
AdS$_{4}$.  For this purpose, the salient features of
AdS$_{4}$ are reviewed in Appendix \ref{appenAdS}.

\subsection{Space-times of the extreme domain walls}

There are three types of extreme domain walls realized
\cite{CGRI,CGI}
in four-dimensional ($N=1$)\ supergravity
theory.  All these walls
have $\beta\equiv 0$.
Thus, they are field theoretic realizations
of the planar $k = q_{0} = 0$ walls.
The space-time metric induced by these walls
is conformally flat with conformal factor
$A(z)$ becoming unity on the M$_{4}$ side of the Type I wall and
falling of as $(\alpha z)^{-2}$ on the AdS$_{4}$ side,
where $\Lambda = -3\alpha^{2}$.
The Type II conformal factor
falls off as $(z\alpha_{1,2})^{-2}$
on the respective sides.
For the Type III wall,
$A(z)$ has an irremovable coordinate singularity
at a finite value of $z$ representing
the affine boundary of the space-time.
In this section we present geodesically complete
extensions of the space-times for the Type I, II, and III extreme
domain walls.  The global space-times of
the Type I wall have been considered previously in Ref.\ \cite{CDGS}
and the Type II wall in Ref.\ \cite{Gibb}.

For each of the walls, we must extend across a
Cauchy horizon on the AdS$_{4}$ side
\cite{CDGS,Gibb}.
The Cauchy horizons occur on the nulls
at $|z|=\infty$ where
$A(z) = 0$,
i.e., where the line element degenerates.
Although these nulls are an infinite
proper distance away, the geodesic distance is
finite.
This type of geometry is familiar from the extreme
black hole space-times \cite{HE,CHAN,Carter1,Carter2}.
The horo-spherical coordinates
must be extended across the Cauchy horizons
on these AdS$_{4}$ sides.
As shown in Appendix \ref{appenAdS},
the need to extend across a Cauchy horizon also
arises in pure AdS$_{4}$.

Cauchy horizons represent the boundary of causal
evolution; therefore, one has the possibility of making
identifications across the
Cauchy horizons which can introduce closed time-like curves (CTCs).
The possibility of CTCs is inherited from the
AdS$_{4}$ portion of the space-time.
Identifications
are especially intriguing
for the type I walls, as CTCs
could lead to supersymmetric
time-machine\cite{FRETAL}
for travellers leaving M$_{4}$ passing across the wall and
then re-emerging into the M$_{4}$ region.
Due to the underlying supersymmetry,
the quantum energy-momentum
infinities \cite{Hawk,Klinkhammer,Visser,Grant} which plague
non-supersymmetric time machines
are avoided\cite{CDGS,Hawk,Burgess}.

There are three possible extensions across the
null Cauchy horizons:
\begin{enumerate}
\item
Move onto a new diamond patch with the scalar field permanently
settled into its vacuum,
i.e., beyond the Cauchy horizon
there is  pure AdS$_{4}$.
\item
In the case of the Type II wall, shift the old diamond along the
null such that the new diamond
is oriented just as the old.  This extension
yields a new wall
as well as a jump in the cosmological constant at the Cauchy
horizon for non-$Z_{2}$ symmetric
walls.\footnote{$Z_{2}$ symmetry means
$\Lambda_{1} = \Lambda_{2}$,
which can be realized for the Type II walls.}
\item
Reflect the old diamond onto the new diamond
across the Cauchy horizon.
This extension leads to a new wall as well as a smooth
matching of the cosmological constant at the horizon.
\end{enumerate}

In the following, we consider geodesic extensions of the third kind.
One reason for doing so is that it yields
the most interesting causal structure for the
resulting space-times.
It is for the third approach that the
causal structure of the Type I and II space-times
exhibit a symmetric
lattice structure similar to those
first realized by the extensions
of Carter for the Kerr and
Reissner-Nordstr{\"{o}}m black holes \cite{Carter1,Carter2}.
The extension
for the Type I wall
realizes the identical causal structure as the
extreme Kerr black hole along its symmetry
axis \cite{HE,CHAN,Carter1}.
Finally, it is through the infinite lattice
for the Type I and II space-times
that one eliminates the
time-like boundary of pure CAdS$_{4}$ (the covering space of
AdS$_{4}$---see Fig.\ \ref{fig9} in Appendix \ref{appenAdS})\
in exchange for a countably infinite
number, $\aleph_{0}$, of isolated points which are an infinite
affine distance away from interior points
(see Figs.\ \ref{fig1} and \ref{fig2}).
For example, the Cauchy problem for
the Type I space-time can be specified by
prescribing initial data on one constant
time slice in an AdS$_{4}$ region
and freely chooosing
boundary data on past null infinity of the
countably infinite
number of adjacent M$_{4}$ spaces (see Fig.\ \ref{fig1}).
In contrast, for pure CAdS$_{4}$, the Cauchy
problem is defined only after prescribing
an infinite amount of boundary
data along the time-like boundaries
which has to be {\em self-consistent\/}
with the specified initial
data \cite{AIS,Breit-Freed}.
The third approach for the extensions can also
be employed in the
case of the non-extreme and ultra-extreme bubbles
as discussed in the next section.
For this case, the infinite lattices are
natural generalizations of the present
extreme space-times.

The three types of extreme space-times,
constructed from the third
kind of geodesic extension
described above,
have the conformal diagrams shown Figs.\ \ref{fig1},
\ref{fig2}, and  \ref{fig3}.
In each of the figures,
the $x$- and $y$-coordinates are suppressed;
therefore, each point represents an infinite plane with distances in the
plane scaled by $A(z)$. The compact null coordinates
$u',v' = 2\tan^{-1}[\alpha(t \mp z)]$ define the axes.
As the figures indicate, these coordinates can be
extended
smoothly
across the
Cauchy horizons (denoted by the dashed nulls)
separating the diamonds
on the AdS$_{4}$ side.
Explicitly this fact is seen by
writing the (1+1)-dimensional line element near the horizon as
$ds^{2} = (\alpha z)^{-2}(dt^{2} - dz^{2}) =
\{ \alpha \sin[1/2(u' - v')] \}^{-2} du' dv'$
which has a smooth extension across the
null $u'=\pi, -\pi < v' < \pi$
as well as all the other Cauchy horizons.
Thus, the null $(u',v')$-coordinates
provide an atlas for describing
the global space-time.
Note that the full (3+1)-dimensional
metric has coordinate
singularites in the $x$- or $y$-directions
crossing the Cauchy horizon.

The extension chosen for the Type I wall \cite{CDGS}
($\sigma_{\text{ext I}} = 2\kappa^{-1}\alpha$)
in Fig.\ \ref{fig1} possesses
the same causal structure as the extreme Kerr black
hole along its symmetry axis \cite{HE,CHAN,Carter1}.
The extension chosen
for the Type II wall in Fig.\ \ref{fig2}
($\sigma_{\text{ext II}} = 2\kappa^{-1}(\alpha_{1} + \alpha_{2})$)
tiles the whole plane with a lattice
of walls.  For the Type III wall
($\sigma_{\text{ext III}} = 2\kappa^{-1}|\alpha_{1} - \alpha_{2}|$)
shown in Fig.\ \ref{fig3},
the conformal factor $A(z)$
diverges
at some finite coordinate $z^{*}$ \cite{CGI}.
This point represents the
boundary of the space-time just as
$z=0$ ($\psi = \pi/2$)
represents the edge of
pure AdS$_{4}$ as seen in
the horo-spherical coordinates (see Fig.\ \ref{fig9}).
As a result, the extension of the Type III wall is causally the same as
pure AdS$_{4}$.

For each of the extensions, the vertices are special
points \cite{Gibb}.
To illustrate their nature, consider a photon moving along one of the
Cauchy horizons, say  $v'= - \pi$.
Since the Cauchy horizon has zero surface gravity, the
photon never reaches the wall in the next
diamond at $u'=\pi, v'=-\pi$~\cite{Boyers}.
These points are an infinite affine distance away from all other
points; i.e., they represent an infinite conformal
compression.
This is  analogous to the situation in
the extreme  Reissner-Nordstr\" om
and Kerr black holes~\cite{HE,CHAN,Carter1,Carter2}.

The conformal diagram of a space-time
containing a Type I or II extreme domain wall centered at  $z=0$
should be compared to that of pure AdS$_{4}$
shown in Fig.\ \ref{fig9}
in Appendix \ref{appenAdS}.
One sees that the time-like affine infinity of AdS$_{4}$ at
$z = 0$ (equivalently $\psi = \pi/2$)
is smoothed out by the wall, which allows
for another space-time region across the boundary of
pure AdS$_{4}$.
In this sense, one can think of the
wall as living at spatial infinity of
AdS$_{4}$.  It has been speculated\cite{Gibb} that the
extreme walls are related to the
{\em ``Membrane at the End of the Universe''\/}
which arises in supermembrane
theory\cite{DUFF}.

\subsection{Space-times of non-extreme and ultra-extreme domain walls}

The generalizations of the extreme walls to $\beta > 0$ yield the non- and
ultra-extreme walls.
As discussed previously, only the $k=\beta^{2}$
version of the line element completely covers the spherical domain
wall bubble.  Again, with $\sin\theta = \beta r$ in Eq.\ (\ref{ansatz}),
the line element for these walls is
\begin{equation}
ds^{2} = A(z)\left( dt^{2} - dz^{2} - \beta^{-2}\cosh^{2}\!\beta t\,
d\Omega_{2}^{2}\right)
\label{lineelement}
\end{equation}
where $d\Omega_{2}^{2} = d\theta^{2} + \sin^{2}\theta d\phi^{2}$
and $A(z)$ is given by one of the $\Lambda = 0$
and $\Lambda = -3\alpha^{2}$ solutions in Eq.\ (\ref{aplus}).
The $\Lambda  = 3\alpha^{2}$ solution is not discussed here.

\subsubsection{Minkowski inside and outside of the non-extreme and
ultra-extreme walls}

The line element for the M$_{4}$ side of the bubble, as
written in the comoving coordinates, is given by
\begin{equation}
   ds^{2} = e^{\mp \beta z}
   \left(
   dt^{2} - dz^{2} - \beta^{-2}\cosh^{2}\!\beta t\,d\Omega_{2}^{2}
   \right).
\label{eq:metric-m4}
\end{equation}
A metric coefficient decreasing away from the wall
represents the
M$_{4}$ side being on the inside of the wall,
and if the exponential is increasing away from it,
the M$_{4}$ side is on the
outside.
For non-extreme walls both sides are on the inside.
For ultra-extreme walls, the side with the highest energy density
(less negative $\Lambda$)\ is on the outside of the wall
and the other side is on the inside.
Recall that we have chosen to orient the $z$-coordinate so that the
most negative $\Lambda$-term is placed on the side $z<0$.
Thus for an ultra-extreme
AdS$_{4}$--M$_{4}$, the M$_{4}$ side is on the outside, and it is described
by the
plus sign solution above, whereas for
an ultra-extreme M$_{4}$--dS$_{4}$, the Minkowski space is on the inside
again described by the plus sign solution but now for $z<0$.

When describing the inside, these coordinates have a
cosmological horizon along the null where $|z|=\infty$.
It is here that the line element
(\ref{eq:metric-m4}) degenerates.
As seen below, this horizon is a Rindler horizon
arising from the hyperbolic motion of the bubble
accelerating away from inertial observers.
For the case of M$_{4}$ being on the
outside  of the bubble, the comoving coordinates
are geodesically complete.  The null where $z=\infty$
represents the usual affine boundary of M$_{4}$
and thus requires no extension.
The two bubbles are complementary
in the sense that the radii of the
spatial region covered by the
respective line elements,
$\underline{r}^{\text{in}}_{\text{out}}
= \beta^{-1} \exp(\mp \beta z) \cosh \!\beta t$,
satisfy:
$0 < \underline{r}^{\text{in}} \le \beta^{-1} \cosh \!\beta t
\le \underline{r}_{ \text{out}} < \infty$.
This complementary nature is also reflected in the
conformal diagrams, as we now discuss.

For the purpose of investigating the
causal structure of the space-time
induced by the
non-extreme and ultra-extreme
bubbles, we introduce
the radial Rindler coordinates
$\underline{t}^{\text{in}}_{\text{out}}
 = \beta^{-1} \exp(\mp \beta z) \sinh\!\beta t$
and
$\underline{r}^{\text{in}}_{\text{out}}
 = \beta^{-1} \exp(\mp \beta z) \cosh\!\beta t$.
This transformation
brings the line element
Eq.\ (\ref{eq:metric-m4})
to the spherically
symmetric form
$ds^{2} = d\underline{t}^{2} -
d\underline{r}^{2} - \underline{r}^{2} d\Omega_{2}^{2}$.
The $(\underline{t}, \underline{r}, \theta, \phi)$ coordinates
define an inertial frame in which the bubbles at $z=0$
live on the hyperbolic trajectory
$\underline{r}^{2} - \underline{t}^{2} =
-\tan(u'/2) \tan(v'/2) = \beta^{-2}$.
Here
$u',v' = 2\tan^{-1}[\beta(\underline{t} \mp \underline{r})]$
are the usual compact null coordinates of
M$_{4}$.
{From} this hyperbolic trajectory, it is apparent that
in order to remain comoving with respect to the
walls, observers require a
constant proper acceleration of
magnitude $\beta$.
Therefore, the comoving frame of the bubble is a
Rindler frame \cite{BirDav}.
The hyperbolic trajectory of the wall is
a general feature of vacuum bubbles,
as discussed in Ref.\ \cite{Coleman,CD}.

For the case where M$_{4}$ is on the inside of the
bubble (e.g.\ the non-extreme case), the wall accelerates
away from inertial observers, and consequently it
produces a Rindler horizon.
This horizon manifests itself as the boundary of the comoving
coordinates $(t,z,\theta,\phi)$.
As this horizon arises from
acceleration,
freely falling particles
reach it with a
finite affine parameter \cite{CGS}.
As a result, an extension of the
comoving coordinates must be provided across the horizon.
As we have globally defined inertial coordinates
$(\underline{t}, \underline{r},\theta,\phi)$,
the unique extension across this horizon onto pure
M$_{4}$ is taken.
On the outside (e.g.\ the M$_{4}$ side of an AdS$_{4}$--M$_{4}$
ultra-extreme domain wall), the bubble
accelerates towards all inertial observers,
and the comoving coordinates
do not require an extension.
Indeed, unless   time-like
observers maintain a constant
proper acceleration not smaller than that of the bubble,
the wall will
eventually  hit them.

The conformal diagram for the
M$_{4}$ side of the
non-extreme bubble is given in
Fig.\ \ref{fig4}.
The dotted line represents the null Rindler
horizons where $|t\pm z|=\infty$.
The unique extension across these horizons is
into pure M$_{4}$ space-time.
As seen by inertial observers in M$_{4}$,
the world tube of the non-extreme bubble is represented by
the surface of the (2+1)-dimensional
de~Sitter hyperboloid. The M$_{4}$
embedding space of dS$_{3}$
is the {\em physical\/} M$_{4}$ space
for this bubble.
Inertial M$_{4}$ observers
are on the inside of the de~Sitter
hyperboloid \cite{Gibb}.
To remain with the wall observers must accelerate towards the wall.
In this
global sense the wall exhibits
``repulsive gravity.''
Fig.\ \ref{fig4} shows a slice of this hyperboloid,
whose surface is the
space-time trajectory of the wall. The bubble's center of
symmetry is the time-like line in the center of the figure.
Opposite points represent spatially antipodal points of the sphere.
The diagram has to be matched to another $q_{0}=\beta^2$ solution on
the wall's  world tube.

In the ultra-extreme AdS$_{4}$--M$_{4}$ case,
the inertial M$_{4}$ observer is
on the outside of the de~Sitter hyperboloid.
The corresponding
conformal diagram (slice of the de~Sitter embedding)\
is given in Fig.\ \ref{fig5}.
Thus, the
M$_{4}$ side of an AdS$_{4}$--M$_{4}$
ultra-extreme wall
is the complement of the
non-extreme M$_{4}$ diagram
of Fig.\ \ref{fig4}.
Here, the nulls are the usual null
infinities of pure M$_{4}$.
The two hyperbolic trajectories
represent the space-time
trajectory of two spatially antipodal points
on the wall surface. Again, the diagram is to be glued to another
solution at the position of the wall, e.g.\ one can put the
inside of the AdS$_{4}$ cylinder (described in the next
section)\ in the hole
where the de~Sitter hyperboloid was taken out,
and identify along the wall's trajectory.

\subsubsection{Anti-de~Sitter inside and outside
of the non- and ultra extreme walls}

The line element for the AdS$_{4}$ side of the bubble,
as written in the comoving coordinates, is given by
\begin{equation}
   ds^{2} = { \beta^{2}
   \over [\alpha \sinh(\beta z - \beta z')]^{2} }
   \left(dt^{2} - dz^{2}
   - \beta^{-2} \cosh^{2} \!\beta t\, d\Omega^{2}_{2} \right),
\label{eq:AdS-wall-co-movingI}
\end{equation}
where
$z>0$
is the side with the less negative cosmological constant.

For the AdS$_{4}$ on the outside of an
ultra-extreme bubble, $z'= z'_{+} > 0$,
which allows $(z-z')$ to vanish at
$z^{*} = z'_{+}$.
$z^{*}$ is an irremovable singularity in the line element
and represents the time-like affine boundary of the
AdS$_{4}$ side of the ultra bubble.
The inside of an ultra-extreme bubble has a more negative
cosmological constant than the outside. This side
as well as the sides with $z<0$ of a non-extreme bubble has $z'= z'_{+}$,
whereas the other inside of a non-extreme bubble with $z>0$ has $z'=z'_{-}$,
and so the line element (\ref{eq:AdS-wall-co-movingI})
degenerates at $|z| = \infty$, which corresponds to the
center(s) of the bubble.
As with the non-extreme M$_{4}$ bubble,
the null on which $|z|=\infty$ represents
a cosmological horizon which is reached
within a finite affine parameter
by time-like and null trajectories \cite{CGS}.
Therefore, an extension across the horizon
is necessary.
This horizon arises from the compactified
hyperbolic
motion of the wall as viewed
in the frame defined by the
Einstein universe coordinates.
This frame is discussed next.

The Einstein universe
coordinates \cite{HE} define a frame
in which the bubbles
exhibit properties
analogous to the M$_{4}$ side
of the bubbles as
seen in inertial M$_{4}$
coordinates.
This frame
provides the
analog of the inertial frame for
the M$_{4}$ case.
The transformation to the Einstein universe frame is
presented in three steps.  Appendix \ref{appencoord}
gives the explicit form of the line element in the
intermediate steps.
(1)\ Define $\ln \Xi = \beta(z-z')$.
{From} the definition of $z'$, it is seen that
$0 < \Xi_{\text{non}} < 1 \le \Xi_{\text{ultra}}$.
(2)\ Introduce the radial Rindler coordinates:
$T = \Xi \sinh\beta t$ and $R = \Xi \cosh\beta t$.
(3)\ Define the compact time-like and radial coordinates:
$T \pm R = \tan[(t_{c} \pm \psi)/2]$.
Performing these transformations on the
line element (\ref{eq:AdS-wall-co-movingI}) yields
$ds^{2} = (\alpha \cos \psi)^{-2}(dt_{c}^{2} -
d\psi^{2} - \sin^{2}\psi d\Omega_{2}^{2})$,
where $-\pi \le t_{c} \pm \psi  \le \pi$ and $0 \le \psi \le \pi/2$.
The spatial center of symmetry is at $\psi = R = 0$.
In the frame defined by the Einstein universe
coordinates, the
bubble at $z=0$ lives on a hyperbolic trajectory
$R^{2} - T^{2} =
-\tan[(t_{c} - \psi) / 2] \tan[(t_{c} + \psi) / 2]
= e^{2\beta z'}$.
To remain stationary with respect to
the bubble---to stay at a fixed
($z,\theta,\phi$)-position---requires
a proper acceleration
of magnitude $|a_{\mu} a^{\mu}| = \alpha^{2} \cosh^{2} \beta(z-z')$.
Non-extreme bubbles accelerate away from observers on both sides,
and thus we expect
a horizon analogous to the
Rindler horizon of the M$_{4}$ side.
The ultra-extreme bubbles approach time-like
observers on the outside and eventually
collide with them. The inside of an ultra-extreme bubble is
indistinguishable from the inside of a non-extreme bubble.

As seen in the Einstein cylinder coordinates,
the conformal diagram
for the AdS$_{4}$ inside of the non-
and ultra-extreme walls
is shown in Fig.\ \ref{fig6}, and the AdS$_{4}$ outside
of an ultra-extreme wall is shown in Fig.\
\ref{fig7}.
Note the complementary nature of the two diagrams;
a relation also exhibited by the M$_{4}$ sides as shown in
Figs.\ \ref{fig4} and \ref{fig5}.
For the non-extreme wall, the space-time is uniquely
extended across the cosmological
horizons (the dotted nulls)\ onto pure AdS$_{4}$
(see Appendix \ref{appenAdS}). The
dashed nulls are the Cauchy horizons of pure AdS$_{4}$.
For the ultra-extreme wall, the region between the
wall at $z=0$ and the time-like boundary of the
space-time, at $z=z^{*} < 0$, is covered by the
$(t,z)$-coordinates.

\subsection{Comments on the extensions}

The previous conformal diagrams represent the minimal extensions
of the space-times on the M$_{4}$ and AdS$_{4}$
sides of the walls.
In the form presented here, they are quite
similar.  The difference
is the
presence of null affine boundaries in the M$_{4}$ regions
as opposed to time-like affine boundaries in the
AdS$_{4}$ regions.
To obtain a complete
space-time, we glue the two sides together at the wall's
 world tube.

Since the
AdS$_{4}$ inside a non-extreme bubble
has a Cauchy horizon,
we are able to introduce a lattice structure
such as discussed in \cite{CGS} and shown in
Fig.\ \ref{fig8}.  This lattice is a generalization of
the extreme lattices shown in Figs.\ \ref{fig1}--\ref{fig3}.
The diagrams of Figs.\ \ref{fig4}--\ref{fig7}
should be thought of as pieces of a space-time
which can be fit together in different ways.
In placing a wall on the AdS$_{4}$ side, we
eliminate a portion of the time-like boundary of
AdS$_{4}$, and just as in the case of the
extreme walls, the Cauchy problem can be formulated
for the infinite lattice space-times
by placing initial
data on one AdS$_{4}$ slice of constant time
and freely specifiable boundary data on a countably
infinite
($\aleph_{0}$)\
number of past null infinities of M$_{4}$ space-times.

These lattices of walls exhibit similarites to the
lattices obtained in extending the non-extreme
black holes \cite{HE,CHAN,Carter1}.
Furthermore, there are local coordinate
relations mapping the horizons in the (time, radial)-directions of the
black hole space-times to the $(t,z)$ horizons in both the
extreme and non-extreme
domain wall systems \cite{CDGS,CGS}.
By making this coordinate connection, it is shown
that near the horizons of both the non-extreme and extreme black holes
(with $d\Omega = 0$), the space-times are locally AdS$_{2}$.

Fig.\ \ref{fig:ultra-tunneling} illustrates the
causal properties of the ultra-extreme bubble corresponding
to the quantum tunneling
event for decay of a meta-stable Minkowski
space-time into the lower energy anti-de~Sitter
vacuum.
At time zero, the bounce becomes the
tunneling two sphere which forms
at rest with a finite radius
and then accelerates along the
hyperbolic Rindler trajectory.
Notice that time-like
trajectories on the M$_{4}$ side
(the meta-stable side)\
eventually collide with the wall and
pass through into the AdS$_{4}$ region
(the stable side).
This situation was addressed
by Coleman and De~Luccia \cite{CD} and later
by Abbott and Coleman \cite{Abb-Col} who
used a singularity theorem of Penrose
\cite{Penrose-theorem} to
conclude that a bubble of AdS$_{4}$ forming
inside an M$_{4}$ region is unstable to the
formation of singularites.
A necessary assumption for establishing this result
is the absence of Cauchy horizons.
In this regard, the dashed null in Fig.\
\ref{fig:ultra-tunneling}
is drawn to
indicate the Cauchy horizon for data placed on
any constant time slice after the bubble forms.
The existence of this horizon
precludes the use of Penrose's theorem
for establishing the singularity result.

\section{Discussion}
\label{discussion}

We have studied the space-times of vacuum
domain walls in general relativity
interpolating between
vacua of non-equal cosmological constant.
Our emphasis has been on vacua of non-positive
cosmological constant since the resulting
solutions are the natural generalizations of
the supersymmetric walls studied in
Refs.\ \cite{CGRI,CGII,CGI,CDGS,Gibb}.
The coordinates in which the
domain wall is
spatially at rest (the comoving coordinates)\
is a useful frame for deducing the local properties of
the solutions through the use of Israel's formalism
of singular hypersurfaces.
We have presented a unified picture of three types of domain walls:
(1)\ extreme (supersymmetric)\
walls which are static, non-compact planar configurations,
with surface energy density $\sigma=\sigma_{\text{ext}}$,
(2)\ non-extreme domain walls
which are bubbles with two insides and energy density
$\sigma=\sigma_{\text{non}} >\sigma_{\text{ext}}$, and
(3)\ the ultra-extreme walls which
correspond to the false vacuum decay bubbles with energy density
$\sigma=\sigma_{\text{ultra}}<\sigma_{\text{ext}}$.

\subsection{Domain walls as gravitational shields}

The extreme AdS$_{4}$--M$_{4}$
space-time (Type I) represents a
solution where the gravitationally repulsive
domain wall exactly compensates the gravitational
attraction caused by the negative energy AdS$_{4}$ vacuum.
This precise balancing is realized with the added feature
of an increased symmetry in the field equations,
i.e.\ the configuration is supersymmetric and
is described by first order rather than
second order equations.
One could say that the Type I
domain wall acts as a gravitational shield
protecting M$_{4}$ observers from the curved
space-time of the AdS$_{4}$ side.

Upon breaking supersymmetry, the energy-density of the
wall can increase or decrease.
Additionally, the non-compact planar geometry of the
wall is replaced by a compact spherical bubble
with time dependent radius.
Increasing the energy of the wall,
one may expect to get a bubble
where the repulsive domain wall overcompensates the
attractive gravity of the AdS$_{4}$ region inside the bubble.
If this were the case, the result would be a
finite object with negative effective gravitational mass.
Negative mass objects introduce very interesting possibilities
\cite{Price}.
For example, controlling  such objects would allow for
a free source of acceleration \cite{Prob}.
This fact can be seen through the equivalence principle,
which implies that the negative mass object
falls towards any positive mass,
which in turn is repelled by the
negative mass object.
Note that this propulsion mechanism
does not violate conservation
of momentum since the negative mass object
has a momentum vector
antiparallel to its velocity.
The present analysis has, however,
shown that for the domain walls considered, space-time is warped so
that observers on both sides of the
would-be negative mass object (the non-extreme bubble)\
are on the inside
of the bubble.  Hence, this kind of negative
mass object {\em cannot\/} be observed from the
outside.

As noted above, these classes of solutions shed light on the nature of
configurations with negative effective
gravitational mass.  They indicate a new kind of censorship,
analogous to Penrose's
cosmic-censorship hypothesis \cite{Censor1,Penrose-censor}.
Nature does not only keep singularities hidden under horizons, but
also seems to prevent us from being outside negative mass objects.
Accordingly, the supersymmetric type I wall system
represents the lowest gravitational energy
state accessible to an outside observer in Minkowski space.
When
the energy-density of the domain wall is further increased, thus
decreasing the effective gravitational mass below zero, the bubble
curves onto both sides making both sides of the bubble
an {\em inside.\/}
We conjecture that similar protection will take place in
all singularity free models and formulate:
\paragraph*{}
{\bf The Positive Mass Conjecture:}
{\em There is no singularity
free solution of Einstein's field equations
for physically acceptable matter sources for which
an exterior observer can see a finite object with a
negative effective gravitational mass.}

By ``physically acceptable'' we mean that matter sources---not
including the vacuum itself---obey
the weak energy condition, and ``singularity free'' does not exclude singular
hypersurfaces tractable by Israel's formalism.

At the same time, we see see that supersymmetry
serves as a {\em positive mass protector\/} in the sense that
supersymmetry is characteristic of
the limiting case where the total effective gravitational mass is zero below
which the domain wall encloses space on both sides.
Again, we note the analogy with cosmic censorship where
supersymmetry has been identified as a cosmic censor \cite{BlackKal}.

\subsection{Relation to physical domain walls}

In this paper, we have presented the
local and global properties of exact solutions of Einstein's field
equations for infinitely thin walls.
To obtain
them it was necessary to assume a high degree of symmetry.
Yet, however idealized these solutions may be,
many of the properties of these solutions
are shared by more realistic, physical domain walls.

Cosmological domain walls are the transition region between two
different vacua and there are
at least three ways of forming walls in a cosmological context:
(1)\ Walls separating vacua that are absolutely stable
against quantum tunneling could form via the Kibble
mechanism \cite{Kibblerev,Vilenkinrev}. (2)\
Primordial walls which could be
born with the universe when it was created
by a quantum tunneling prosess out of nothing
\cite{QuantumCosmology,HartleHawking,Linde}.  This
tunneling can yield regions of
different vacuum states separated by walls.
$(3)$ Walls could also be the boundary of
a bubble created from a first order phase transition between
a false and true vacuum through the mechanism of quantum
tunneling\cite{Coleman,CD}.

The stable vacuum manifold for a scalar field theory
consists of minima of the scalar potential
which are not connected by quantum tunneling.
For scalar theories without
gravity, any potential possessing minima of the
same energy are degenerate; there is no tunneling because the
corresponding bounce instanton has infinite action \cite{Coleman}.
Conversely, if there is a non-zero difference in the value of the scalar
energy at the minima, there will be a finite probability for
decay to the global minima
through the formation of a tunneling bubble\cite{Coleman,Fram,COL,CALL}.
In the case with gravity,
the coupling of a constant
scalar vacuum energy to gravity through the
{\em effective\/}
cosmological constant, the value,
and most importantly the {\em sign,\/}
of the vacuum energy are essential for understanding the
stability of a vacuum to quantum decay.

For dS$_{4}$,
gravity always tends to increase the probability for decay
to a lower cosmological constant vacuum relative to the
corresponding process in the non-gravitational theory \cite{CD}.
Therefore, the domain
walls considered in
Sec.\ \ref{grsection}
separating different
cosmological constant vacua, where at least one vacuum is dS$_{4}$,
represent the local properties for the
classical evolution of a bubble formed
from such a first order phase transition.
Discussions of these bubbles can be found in the literature
\cite{BKT,SATO,GUTHETAL,BEREZIN}.
Only those walls separating
dS$_{4}$ of the {\em same\/} cosmological constant
are stable topological walls,
and these are unstable to perturbations in the adjacent
vacuum energy.

If one vacuum is M$_{4}$ and the other AdS$_{4}$ \cite{CD}, or if both
vacua are AdS$_{4}$ \cite{CGRII},
there is a lowering of the probability for decay
relative to the corresponding non-gravitational case.
If the scalar potential energy barrier is such that
the surface energy of the tunneling bubble
$\sigma_{{\rm bubble}}$ is greater than
or equal  to $2\kappa^{-1}(\alpha_{1} + \alpha_{2})$, where
$\Lambda_{1} = -3\alpha^{2}_{1}$ and
$\Lambda_{2} = -3\alpha^{2}_{2}$,
then the tunneling instanton
has infinite action which renders the
tunneling probability to zero.
As a result, stable
topological domain walls can exist between such vacua even
without the scalar energy in the two vacua being the same.

We have discussed three kinds of walls:
the non-extreme,
extreme, and ultra-extreme walls.
Here we point out the relation between the
idealized models and physical domain walls.
The non-extreme wall,
whose surface energy is
$\sigma_{\text{non}} > 2\kappa^{-1}(\alpha_{1} + \alpha_{2})$,
correspond to walls
which may have formed by the Kibble mechanism or
to primordial domain walls born with the universe at the time of
creation.
If one starts out with a universe with closed spatial sections,
and allows
the scalar field to
fall into two different vacua, these
two regions
can be separated by a non-extreme bubble with two insides.
A lower dimensional picture will
illustrate that this can happen.
Assume that we live in a two dimensional
space, represented by the surface of a sphere.
Let us divide it in two halves
along a great circle. Both
sides are on the inside of the
great circle.
Therefore,
if the initial spatial geometry is closed,
the seemingly strange topology of the non-extreme
wall causes no problem to the creation of walls of this type through the
Kibble mechanism.
Indeed, it has been speculated \cite{BKT,GUTHETAL}
that we actually live inside
a non-extreme bubble.

The ultra-extreme wall,
whose surface energy is
$\sigma_{\text{ultra}} < 2\kappa^{-1}(\alpha_{1} + \alpha_{2})$,
represents
the classical evolution of a quantum tunneling bubble
(which also could be primordial)\ that forms when
a metastable region of Minkowski or anti-de~Sitter false
vacuum decays into a lower energy anti-de~Sitter vacuum.
First order phase transitions of this type can
occur in theories for which the potential energy barrier
is insufficient to suppress the tunneling.

There is no
tunneling \cite{CGRII} between supersymmetric vacua since the
minimal energy tunneling bubble saturates the
Coleman-De~Luccia bound:
$\sigma_{{\text{bubble}}}=
2\kappa^{-1}(\alpha_{1} + \alpha_{2})$, thus rendering
the decay probability to zero.
This result is exact to all orders in Newton's constant and applies
to both supersymmetric AdS$_{4}$--AdS$_{4}$ and AdS$_{4}$--M$_{4}$ vacua.
As a consequence, all supersymmetric vacua are
degenerate in the sense that
there is no tunneling between them.
The non-compact planar extreme walls
with surface tension
$\tau=\sigma_{{\text{wall}}}=
2\kappa^{-1}(\alpha_{1} + \alpha_{2})$
interpolate between these supersymmetric
vacua~\cite{CGRI,CGII,CGI}.
The extreme walls separate
supersymmetric vacua of non-positive
cosmological constant where at least one of the
vacua is anti-de~Sitter and they are the
configurations
intermediate to
the spherical non- and
ultra-extreme bubbles.

\subsection{Final remarks}

We have given a unified, global presentation of the gravitational aspects
of domain walls.
Apart from the theoretical insight in
General Relativity and Supersymmetry gained from these walls,
and the relation to possible
physical realizations of domain walls in Nature, we also would like to
point out the didactic value of these solutions. In this very simple model
one encounters maximally symmetric spaces, Tolman's mass,
Israel's thin wall formalism,
comoving coordinates and the FLRW metric, the de~Sitter hyperboloid,
Rindler motion, cosmological horizons, Cauchy horizons, and the problem
of geodesic incompleteness. For this reason,
it is an ideal tool for illustrating
many important concepts and techniques in General Relativity.

The r{\^{o}}le of the
supersymmetric domain wall as a perfectly balanced planar configuration
{\em intermediate\/} between two types of spherically symmetric bubbles have
been explained. For wall energy densities below
the supersymmetric value (ultra-extreme),
the wall surface curves away from time-like observers
on the side with the highest
vacuum energy density and accelerates towards them.
If the wall energy density is above the
supersymmetric value (non-extreme),
the wall curves towards observers
on both sides and accelerates away from them.
Analysis of these non-extreme bubbles has also yielded
the ``positive mass conjecture,''
which precludes the free acceleration
realized from gravitating bodies of negative
effective mass. We also note that supersymmetry serves as a
``positive mass protector.''

This study has provided the first steps towards a
theoretical foundation
for studying the cosmological effects of
these walls; especially those arising
from supergravity where the vacua are either
supersymmetric or have spontaneously broken
supersymmetry. It should, however, be emphasized that
physically realistic
domain walls break many of our symmetry assumptions by being
wiggly and  non-isotropic.

To conclude, the domain wall solutions give
valuable insight in non-perturbative
aspects of gravity.
In particular, it is noteworthy that
gravity, however weak it might be, determines
the topology of the domain walls. In this way, non-perturbative
gravitational effects play a very important r{\^{o}}le
both in the evolution of cosmic domain walls and
in the evolution of the inside of quantum tunneling bubbles.

\section*{Acknowledgments}
We would like to thank J.\ Garriga, A.\ D.\ Linde, P.\ J.\ Steinhardt, and
A.\ Vilenkin for discussions.
H. H. Soleng would like to thank Professor Paul J. Steinhardt
for the invitation to spend the sabbatical year 1992-93 at the
University of Pennsylvania, and he would like to express
his gratitude to the Department of Physics
for its
hospitality.
This work was supported
in part by  U. S. DOE Grant No.\ DOE-EY-76-C-02-3071,
NATO Research Grant No.\ 900-700 (M. C.),
Fridtjof Nansen Foundation Grant No.\
152/92 (H. H. S.),
Lise and Arnfinn Heje's Foundation
Ref.\ No.\
0F0377/92
(H. H. S.), and by the Norwegian Research Council for Science and the
Humanities (NAVF), Grant No.\ 420.92/022 (H. H. S.).

\appendix
\section{Coordinate transformations}
\label{appencoord}

\renewcommand{\theequation}{\Alph{section}.\arabic{equation}}

In this paper, we produced a number of line elements describing
vacuum solutions external to domain walls.
In this appendix we provide local transformations
relating the coordinates natural for the wall geometry
to canonical coordinates for the appropriate vacuum solutions.
We present results only for the cases $q_{0} = 0, ~+\beta^{2}$,
since the case $q_{0} = -\beta^{2}$
corresponds to a solution with a metric that is singular for
certain proper times.

\subsection{Extremal walls ($q_{0} = 0$)}

The $k=0$  solutions are written
in the canonical static coordinates:
Cartesian for M$_{4}$ and horo-spherical
for AdS$_{4}$.  The horo-spherical coordinates
are discussed in Refs.\ \cite{Gibb,Griffies}.

The $k=-\beta^{2}$ line elements are related to the
$k=0$ line elements through a coordinate transformation
not involving the spatial coordinate $z$.
For this case, consider the metric for flat (2+1)-dimensional space-time
written in planar polar coordinates
\begin{equation}
ds^{2} = d\underline{t}^{2} - d\underline{r}^{2} -
\underline{r}^{2} d\phi^{2} .
\end{equation}
Introducing  the Milne coordinates~\cite{Robertson,BirDav}
\begin{equation}
\left. \begin{array}{rcl}
\underline{t} &=&
t \cosh\!\beta \chi\\
\underline{r} &=&
t \sinh\!\beta \chi,
       \end{array}
\right.
\label{milnecoord}
\end{equation}
brings the line element to
\begin{equation}
ds^{2} = dt^{2} -
(\beta t)^{2}\left[ d\chi^{2}
+ \beta^{-2} \sinh^{2}\!\beta \chi\,d\phi^{2}\right].
\label{milneII}
\end{equation}
Transforming to a radial coordinate $ r = \beta^{-1} \sinh\!\beta \chi$
yields
\begin{equation}
ds^{2} = dt^{2} -
(\beta t)^{2}\left( { dr^{2} \over 1
+ (\beta r)^{2} } + r^{2} d\phi^{2} \right).
\label{milneIII}
\end{equation}
Finally, addding an extra dimension coordinated by $z$ and allowing
for
a non-positive cosmological constant yields
\begin{equation}
ds^{2} = A(z) \left[ dt^{2} - dz^{2} -
(\beta t)^{2}\left( { dr^{2} \over 1
+ (\beta r)^{2} } + r^{2} d\phi^{2} \right) \right],
\label{milneIV}
\end{equation}
where
$A(z)$ is one of the $q_{0}=0$ (extremal) solutions
in Eqs.\ (\ref{anull}).

\subsection{Non-  and ultra-extreme walls ($q_{0} = \beta^{2}$)}

\subsubsection{The $\Lambda = -3\alpha^{2}$ solution}

Consider the AdS$_{4}$ metric as written in the Einstein universe coordinates
discussed in Appendix \ref{appenAdS}
\begin{equation}
   ds^{2} = \left(\alpha \cos\psi\right)^{-2} \left( dt_{c}^{2} - d\psi^{2}
   - \sin^{2}\!\psi \,d\Omega^{2}_{2} \right).
\label{eq:AdS-Einstein}
\end{equation}
Note that these coordinates,
just as the
Schwarzschild-coordinates,
cover the whole AdS$_{4}$ manifold.
In the latter coordinates the metric is
\begin{equation}
  \alpha^{2} ds^{2} = \left( 1+\rho^2 \right)dt_{c}^{2} -
   \left( 1+\rho^2 \right)^{-1}d\rho^2-\rho^2d\Omega^2_{2} ,
\label{eq:AdS-static}
\end{equation}
where $\rho = \tan \psi$.
We now decompactify the coordinates $(t_{c}, \psi)$
by introducing
\begin{equation}
   T \pm R = \tan\left[\frac{1}{2}(t_{c} \pm \psi)\right].
\label{eq:decompactify}
\end{equation}
This transformation, restricted to the branch
$-\pi \le t_{c} \pm \psi \le \pi$,
brings the line element
(\ref{eq:AdS-Einstein}) to
\begin{equation}
   ds^{2} = { 4 \over \alpha^{2} (  T^{2} - R^{2}+1 )^{2} }
            \left( dT^{2} - dR^{2} - R^{2} d\Omega^{2}_{2} \right).
\label{eq:AdS-intermed}
\end{equation}
Transforming to the {\em radial\/} Rindler coordinates \cite{BirDav}
$T= e^{\beta(z-z')} \sinh \!\beta t$
and
$R= e^{\beta(z-z')} \cosh\! \beta t$,
where $z'$ is given
in Eq.\ (\ref{deltadet}),
brings (\ref{eq:AdS-intermed})
to
\begin{equation}
   ds^{2} = { \beta^{2}
   \over \left[\alpha \sinh(\beta z - \beta z')\right]^{2} }
   \left(dt^{2} - dz^{2}
    - \beta^{-2} \cosh^{2} \!\beta t\, d\Omega^{2}_{2} \right).
\label{eq:AdS-wall-co-moving}
\end{equation}
This metric corresponds to the first of the solutions
of Eq.\ (\ref{aplus}).

\subsubsection{The $\Lambda = 0$ solution}

Starting from the flat space-time metric written in
spherical coordinates,
\begin{equation}
ds^{2} = d\underline{t}^{2} -
d\underline{r}^{2} - \underline{r}^{2} d\Omega^{2}_{2}\; ,
\end{equation}
and transforming to the {\em radial\/} Rindler coordinates~\cite{BirDav}
\begin{equation}
\left.\begin{array}{ccl}
\underline{t} &=& \beta^{-1} e^{ \pm \beta z } \sinh\!\beta t \\
\underline{r} &=& \beta^{-1} e^{ \pm \beta z } \cosh\!\beta t
      \end{array}
\right.
\label{radialrindler}
\end{equation}
brings the line element to the desired form
\begin{equation}
   ds^{2} = e^{ \pm 2 \beta z }\left( dt^{2} - dz^{2} -
   \beta^{-2} \cosh^{2}\!\beta t\, d\Omega^{2}_{2} \right).
\label{desiredform}
\end{equation}
This metric corresponds to the second of the solutions (\ref{aplus}).

\subsubsection{The $\Lambda = 3\alpha^{2}$ solution}

Starting from the de~Sitter metric in canonical coordinates
\begin{equation}
   ds^2=d\overline{T}^2-\cosh^2\! \alpha \overline{T}
   \left(
   \frac{d\rho^2}{1-\alpha^2\rho^2}
   +\rho^2 d\Omega^{2}_{2}
   \right)
\label{canonical-ds}
\end{equation}
and defining
$(\cos t_{c})^{-1} \equiv \cosh \!\alpha \overline{T}$
and $\sin\!\psi\equiv\alpha\rho$ gives the
metric
\begin{equation}
   ds^2=\frac{1}{\alpha^2\cos^2 t_{c}}\left(dt^{2}_{c}-d\psi^2
   -\sin^2\!\psi\,d\Omega^2_{2}\right).
\end{equation}
We now decompactify the coordinates $(t_{c},\psi)$ by defining
$T$ and $R$ as in Eq.\ (\ref{eq:decompactify}).
This transformation restricted to $-\pi\leq t_{c}\pm\psi\leq\pi$,
brings the de~Sitter metric (\ref{canonical-ds})\ to the form
\begin{equation}
   ds^2=\frac{4}{\alpha^2(T^2-R^2-1)^2}\left(dT^2
   -dR^2-R^2d\Omega^{2}_{2}\right).
\label{intermed-ds}
\end{equation}
Transforming to the Rindler coordinates
$T= e^{\beta(z-z'')} \sinh\! \beta t$
and
$R= e^{\beta(z-z'')} \cosh\! \beta t$,
where $z''$
is given
in Eq.\ (\ref{gammadet}),
brings (\ref{intermed-ds})
to
\begin{equation}
   ds^{2} = { \beta^{2}
   \over \left[\alpha \cosh(\beta z - \beta z'')\right]^{2} }
   \left(dt^{2} - dz^{2} -
   \beta^{-2} \cosh^{2} \!\beta t\, d\Omega^{2}_{2} \right).
\label{eq:dS-wall-co-moving}
\end{equation}
This metric corresponds to the third of the
solutions in Eq.\ (\ref{aplus}).

\section{Aspects of AdS}  \label{appenAdS}

To help in understanding the properties of space-times
with a domain wall where at least one side is AdS$_{4}$,
we present here the salient features of
AdS$_{4}$.
More detailed discussions of AdS$_{4}$
can be found in
Hawking and Ellis \cite{HE}, Avis et al.\ \cite{AIS},
Gibbons~\cite{Gibb}, and Griffies \cite{Griffies}.

\subsection{AdS$_{4}$}

AdS$_{4}$ is the maximally
symmetric solution to Einstein gravity
in four space-time dimensions with a negative cosmological
constant $\Lambda\equiv -3\alpha^2$
\begin{equation}
   {\cal R}_{\mu \nu} - {1 \over 2}{\cal R}g_{\mu \nu}
   = -3\alpha^{2}g_{\mu \nu}.
\label{einstein}
\end{equation}
The length scale of AdS is set by its
``radius'' $\alpha^{-1}$.

AdS$_{4}$ is isomorphic
to the coset $SO(3,2)/SO(3,1)$ and is realized
as a hyperboloid embedded in a flat five dimensional
space with {\em two\/} time-like directions:
$\eta_{AB}Y^{A}Y^{B} = \alpha^{-2}$ and
$\eta_{AB} = {\mbox{diag}}(+1, -1, -1, -1, +1)$.
One can completely cover the hyperboloid and thus all
of AdS$_{4}$ with the following choice of {\em spherical\/} or
{\em Einstein universe\/} coordinates:
\begin{eqnarray}
   \alpha Y^{0} &=& \cos t_{c}  \sec\psi   \nonumber \\
   \alpha Y^{1} &=& \sin\theta \cos\phi  \tan\psi  \nonumber \\
   \alpha Y^{2} &=& \sin\theta \sin\phi  \tan\psi  \nonumber \\
   \alpha Y^{3} &=& \cos\theta \tan\psi    \nonumber \\
   \alpha Y^{4} &=& - \sin t_{c} \sec\psi.
\label{AppembedI}
\end{eqnarray}
The line element
$ds^{2} = \eta_{AB}dY^{A}dY^{B}$ takes on the form
\begin{equation}
   ds^{2} = \left(\alpha \cos\psi\right)^{-2} \left( dt_{c}^{2} - d\psi^{2}
   - \sin^{2}\!\psi\, d\Omega^{2}_{2} \right) ,
\label{AppmetricI}
\end{equation}
and the range on these
dimensionless
coordinates sufficient to cover all
of AdS$_{4}$ is $-\pi \le t_{c} \le \pi, \ \ 0 \le \psi < \pi/2,
\ \ 0 \le \theta \le  \pi, \ \ 0 \le \phi < 2\pi$.
The line element without the $(\alpha \cos \psi)^{-2}$ conformal
factor is that of the static Einstein universe, where in addition
$0 \le \psi \le \pi$.
The spherical coordinates are inextendible;  i.e.,
all geodesic motion on AdS$_{4}$ is described in these
coordinates.

AdS$_{4}$ has the topology ${\bf S}^{1}$(time)$\times {\bf R}^{3}$(space)
which means it has a periodic time and
closed time-like curves (CTCs).
This fact is intimately related to the negative vacuum energy density
(negative cosmological constant) which violates the familiar
positive energy conditions.
In the context of Type II walls, Gibbons discusses this point in
Ref.\ \cite{Gibb}.
If satisfied, these energy conditions
restrict space-times to have a non-periodic time-like coordinate.
It is possible to avoid CTCs
by using the covering space-time
CAdS$_{4}$ in which the
compact time coordinate $t_{c}$ is allowed to range over the
whole real line.  In other words,
we unwrap the ${\bf S}^{1}$(time)
rendering the topology ${\bf R}^{4}$.
Nevertheless, neither CAdS$_{4}$ nor AdS$_{4}$ have a
Cauchy surface, and on account of this, neither AdS$_{4}$
nor CAdS$_{4}$ is
globally hyperbolic.  Therefore,
an infinite amount of
boundary data as well as initial data
must be introduced
in order to properly
define the Cauchy problem~\cite{AIS,Breit-Freed}.
AdS$_{4}$ can be regarded as the canonical space-time in which the
issues of Cauchy horizons and CTCs arise.

\subsection{AdS$_{4}$ mapped on the Einstein Universe}

In understanding the global aspects of AdS$_{4}$, such as its causal
and geodesic properties,
it is useful to map AdS$_{4}$ onto the Einstein cylinder.
The static Einstein universe has the
topology ${\bf R}$(time)$\times {\bf S}^{3}$(space)
and the line element
\begin{equation}
ds^{2}_{c} =  dt_{c}^{2} - d\psi^{2}
          - \sin^{2}\!\psi \, d\Omega^{2}_{2} =
dt_{c}^{2} - d\Omega^{2}_{3}
\label{metricEinstein}
\end{equation}
where
$-\infty < t_{c} < \infty, \ \ 0 \le  \psi, \theta \le \pi, \ \
0 \le  \phi < 2\pi$.
CAdS$_{4}$ is conformal to the half
of the static Einstein universe
with $0 \le \psi <  \pi/2$.
If we suppress the $S^{2}$ coordinates $\theta$ and $\phi$, the
Einstein universe is a cylinder with $\psi$ running along the
${\bf S}^{1}$ and $t_{c}$ along the $\bf{R}$.  Since $0 \le \psi <  \pi$,
opposite sides of the cylinder are identified.
Cutting the cylinder at $\psi = \pi$ yields Fig.\ \ref{fig9}.
AdS$_{4}$ is the
region where $0 \le \psi < \pi/2$, $-\pi \le t_{c} \le \pi$
and each point on the cylinder a two-sphere ${\bf S}^2$ of radius
$\alpha^{-1} \tan\psi$. Radial nulls are at $45^{\circ}$.
The non-trivial
causal structure of AdS$_{d}$, where $d \ge 2$,
is understood from investigating AdS$_{2}$.

\subsection{Geodesic structure of AdS$_{2}$ as seen on the Einstein cylinder}

Integration of the geodesic equations on AdS$_{d}$ is
facilitated by the maximal symmetry of the space-time.
For AdS$_{4}$,
maximal symmetry allows us to lay down coordinates
such that the geodesic of interest has zero
$\theta$ and $\phi$ angular momentum.
The geodesics are thus determined by the AdS$_{2}$
line element
\begin{equation}
   ds^{2} = (\alpha \cos \psi)^{-2}\left(dt_{c}^{2} - d\psi^{2}\right).
\label{sphericaltwo}
\end{equation}
Introducing the conserved
energy parameter\footnote{The mass dimension of $E$ is $-1$.}
$E = (\alpha \cos\psi)^{-2} dt_{c}/d\tau$,
where $\tau$ is the
affine parameter
(proper time for the time-like geodesics),
allows us to express the
line element for
these geodesics $ds^{2} = d\tau^{2} > 0$ as
\begin{equation}
   \left(d\psi / d\tau\right)^{2} =
   \left(\alpha \cos\psi\right)^{4}
   \left[ E^{2} - (\cos\psi)^{-2}\right].
\label{geodesicI}
\end{equation}
Note that the time-like nature of the
motion imposes the constraint
$(\alpha E)^{2} \ge 1$.
Solving for $\psi(\tau)$
and $t_{c}(\tau)$ is straightforward.
Eliminating $\tau$
yields the periodic world-line for the time-like
test particles
\begin{equation}
   \sin^{2} \psi =
   \left[1 - (\alpha E)^{-2}\right]
   \sin^{2}\! t_{c}.
\label{geodesicIV}
\end{equation}
Setting $E$ to infinity in (\ref{geodesicIV}),
yields the null world-line
\begin{equation}
   \sin^{2}\!\psi = \sin^{2}\! t_{c}.
\label{geodesicV}
\end{equation}
Finite energy
time-like geodesics never reach
$\psi = \pi/2$.  Rather,
the constant curvature of AdS$_{4}$ acts as a perfect gravitational
harmonic oscillator causing the time-like test particles
to always return to their original position in a proper time
of  $\pi/\alpha$.  Thus, the proper time
period for a full transit through a fundamental
AdS$_{4}$ domain is $2\pi/\alpha$. This oscillatory motion
is consistent with AdS$_{4}$ having the positive gravitational
mass density of $6\kappa^{-1}\alpha^{2}$
at every point; cf.\ Eq.\ (\ref{adsenergi}).
These world-lines are indicated in Fig.\ \ref{fig9}.

\subsection{Cauchy horizon for AdS$_{2}$}

As can be seen from obtaining the
null geodesics in terms of the null
affine parameter \cite{Griffies},
nulls reach $\psi = \pi/2$
only after an infinite affine parameter.
This means that $\psi = \pi/2$ is
identified as the affine boundary of
AdS$_{d}$.
It is here that the line element
(\ref{AppmetricI}) has an irremovable
coordinate singularity
which is characterictic of affine boundaries.
The time-like nature of $\psi = \pi/2$
precludes AdS$_{4}$ from having a Cauchy surface.
Space-times with a Cauchy surface allow for
a deterministic description
of the classical evolution of free fields
given a sufficient amount of initial data
placed on a space-like slice~\cite{HE}.
Equivalently, every point to the
future of a Cauchy surface
must have a past directed light-cone
which intersects it.
It follows that a Cauchy surface cannot
be time-like.
Since spatial infinity in AdS$_{4}$ is time-like,
information evolved from some initial space-like
slice, say $t_{c} = -\pi$ in Fig.\ \ref{fig9},
can be corrupted from data flowing
in from beyond the null diamonds.
These nulls represent the Cauchy horizon for
this data.





\begin{figure}
\caption{Conformal diagram of the extreme Type I domain wall, which
   separates AdS$_{4}$ from  M$_{4}$.
   The $x$- and $y$-directions are suppressed; therefore,
   each point represents an infinite plane with distances in the
   plane conformally compressed by $A(z)$.
   The compact null coordinates are $u',v' = 2\tan^{-1}[\alpha(t \mp z)]$.
   The domain wall is the double time-like arc splitting the diamonds.
   The complete extension consists of an infinite lattice of diamonds.
   The vertices are infinitely conformally compressed
   points; i.e., they are an infinite affine distance
   away from points interior.
   Cauchy horizons for data placed on the constant time slices
   in one diamond are the dashed nulls separating the AdS$_{4}$  patches.
   The walls smooth out the singularities at the time-like boundaries
   of pure AdS$_{4}$ seen in Fig.\ 10.
   The removal of the time-like boundary allows for
   a formulation of the Cauchy problem on the covering space-time
   which prescribes
   initial data on one slice across an AdS$_{4}$ region
   and freely chooses boundary data on the past null
   infinities of the countably infinite number
   of M$_{4}$ regions.
   Note the similarity of the extension taken here to
   that of the extreme
   Kerr black hole along its symmetry axis [30,32]
   (Diagram taken after Ref.\ [18]).}
\label{fig1}
\end{figure}

\begin{figure}
\caption{Conformal diagram of the extreme Type II domain wall.
   Conventions follow Fig.\ 1.
   AdS$_{4}$ regions are on both sides of
   the wall. As there are Cauchy horizons on both sides
   of the wall, the geodesically complete extension
   covers the whole plane with an infinite
   lattice of domain wall diamonds.}
\label{fig2}
\end{figure}

\begin{figure}
\caption{Conformal diagram of the extreme Type III domain wall.
   Conventions follow Fig.\ 1.
   AdS$_{4}$ regions are on both sides of the domain wall.
   The irremovable singularity at $z=z^{*}=z'_{-}$
   is represented by the time-like affine
   boundaries.  This diagram has the same causal structure
   as pure AdS$_{4}$ seen in Fig.\ 10
   (see also Ref.\ [20]).
   In this way, this system can be thought of as a generalized
   AdS$_{4}$.}
\label{fig3}
\end{figure}

\begin{figure}
\caption{Conformal diagram for the
   M$_{4}$ side of the non-extreme bubble
   or the M$_{4}$ side inside an ultra-extreme bubble.
   This diagram is
   part of (3+1)-dimensional Minkowski space
   as seen in the compactified
   $(\underline{t},\underline{r})$
   plane.  Angular coordinates $(\theta,\phi)$ are suppressed.
   The axis of symmetry represents the world
   line of the center of the bubble at $\underline{r} = 0$.
   Opposite points on the right and left sides of
   $\underline{r} = 0$ represent antipodal points
   $\theta \rightarrow \pi - \theta$ and
   $\phi \rightarrow \phi + \pi$.
   The time direction increases upward.
   The solid curved lines asymptoting to the
   dotted nulls are the world-lines of anti-podal
   points of the non-extreme bubble wall
   at $z=0$, or equivalently
   $\underline{r}^{2} - \underline{t}^{2} =
   -\tan(u'/2) \tan(v'/2) = \beta^{-2}$.
   This diagram is a cross-section
   of the hyperboloid of dS$_{3}$ as
   embedded in M$_{4}$
   (see Ref.\ [30]
   for the analogous case of dS$_{4}$).
   The rest frame of the wall is a
   Rindler frame whose acceleration has
   magnitude $\beta$.
   The dotted nulls are the Rindler horizons
   on which the comoving coordinates $(t,z)$
   degenerate ($z = \infty$).
   In order to remain with the wall, observers
   must accelerate towards it.
   In this sense, the non-extreme bubble exhibits
   ``repulsinve gravity.''
   The unique extension of the
   comoving coordinates across the Rindler horizons
   is onto pure Minkowski space-time.}
\label{fig4}
\end{figure}

\begin{figure}
\caption{Conformal diagram for the
   M$_{4}$ region outside
   of the AdS$_{4}$--M$_{4}$ ultra-extreme bubble.
   This diagram is the complement of the
   non-extreme bubble of Fig.\ 4.
   The two sides represent spatially
   anti-podal pieces of the spherically
   symmetric space-time.
   In this case, the M$_{4}$ side
   corresponds to the outside of the de~Sitter hyperboloid
   of Fig.\ 4.
   These wedges are covered by
   the comoving coordinates $(t,z)$.
   The solid curved line is
   the hyperbolic trajectory of the
   wall at $z=0$.
   The solid nulls
   are the affine boundaries.
   Time-like observers
   with insufficient acceleration eventually encounter the wall.
   In this sense, the ultra-extreme bubble exhibits ``attractive gravity.''
   }
\label{fig5}
\end{figure}

\begin{figure}
\caption{Conformal diagram for an
   AdS$_{4}$ region inside the non-extreme or
   ultra-extreme bubble.
   This diagram is part of pure AdS$_{4}$ as seen in the
   Einstein cylinder coordinates $(t_{c}, \psi)$
   (see Fig.\ 10).
   The angular coordinates $(\theta,\phi)$ are suppressed, and
   the center of symmetry represents the world-line
   of the center of the bubble at
   $R = \psi = 0$.  The vertical boundaries
   are the time-like boundaries of pure AdS$_{4}$
   at $\psi = \pi/2$ or $R=\infty$.
   Opposite points on the right and left of
   $\psi=0$ represent antipodal points
   $\theta \rightarrow \pi - \theta$ and
   $\phi \rightarrow \phi + \pi$.
   The time direction increases upward.
   The solid curved lines asymptoting to the
   dotted nulls are the world-lines of anti-podal
   points of the non-extreme bubble wall
   at $z=0$, or equivalently $R^{2} - T^{2} =
   -\tan[(t_{c} - \psi) / 2] \tan[(t_{c} + \psi) / 2]
   = \exp(2\beta z')$.
   The wall on the AdS$_{4}$ side sweeps
   out a compactified hyperbolic trajectory
   over half the fundamental domain of pure AdS$_{4}$
   (e.g.\ $-\pi \le t_{c} \le 0$ in Fig.\ 10).
   The rest frame of the wall is a
   frame whose acceleration has
   magnitude $\alpha \cosh \beta(z-z')$.
   The dotted nulls are the horizons
   on which the comoving coordinates $(t,z)$
   degenerate ($z = -\infty$).
   The unique extension of the
   comoving coordinates across the horizons
   is onto pure AdS$_{4}$.
   The dashed nulls represent the Cauchy horizons of
   pure AdS$_{4}$.}
\label{fig6}
\end{figure}

\begin{figure}
\caption{Conformal diagram for the
   AdS$_{4}$ region outside of the ultra-extreme bubble.
   This diagram is the complement of the non-extreme bubble
   Fig.\ 6.
   Conventions are as in that figure.
   The comoving coordinates $(t,z)$ cover these
   slivers.  The vertical line is the affine boundary
   and the curved line is
   the trajectory of the bubble wall.
   Time-like trajectories in these patches
   eventually encounter
   the wall, just as for the M$_{4}$ ultra-extreme bubble.}
\label{fig7}
\end{figure}

\begin{figure}
\caption{Conformal diagram for the
   gluing together of bubble insides
   to form a lattice of non-extreme bubbles.
   The left diagram is that of the
   AdS$_{4}$ side shown in Fig.\ 6
   with two successive wall
   trajectories asymptoting to the time-like
   affine boundaries.
   For the infinite lattice,
   there are an infinite number of these walls.
   The right diagram is that of the
   adjoining non-extreme
   M$_{4}$ bubbles shown in Fig.\ 4.
   These regions are not overlapping,
   i.e., they do not touch.
   Movement between the M$_{4}$ patches
   is only realized by first passing
   through the bubble, into the
   AdS$_{4}$ side, and then back out the next bubble.
   Periodic identifications lead to CTCs
   for the space-time.
   The two diagrams are identified across the
   wall region by revolving and rotating
   the de~Sitter hyperboloid of the wall space-time
   as embedded in Minkowski space (the M$_{4}$ side)\
   around the AdS$_{4}$ cylinder
   and identifying adjacent points of the two wall
   regions.  In a similar manner to the extreme lattices
   of Figs.\ 1 and 2, the replacement of the time-like
   affine boundary allows for a more conventional
   Cauchy problem where initial data is placed
   a time-like slice across an AdS$_{4}$ region and
   boundary data is given
   on the past null infinities of the M$_{4}$ sides.}
\label{fig8}
\end{figure}

\begin{figure}
\caption{Conformal diagram for the
   classical evolution of a quantum
   tunneling event where M$_{4}$ ({\bf M})
   decays into AdS$_{4}$ ({\bf A}).
   This is the ultra-extreme bubble where
   M$_{4}$ is outside and the lower
   cosmological constant (and thus higher pressure)\
   AdS$_{4}$ region is inside.
   The gluing of the diagrams
   is performed as in
   Fig.\  8.
   The quantum tunneling bubble forms at
   real time
   ($t=\underline{t} = t_{c}$)
   $=$ imaginary time $=$ zero.
   The tunneling event
   cannot be described by classical gravity,
   thus the jagged line in this region.
}
\label{fig:ultra-tunneling}
\end{figure}

\begin{figure}
\caption{AdS$_{4}$ as seen on the Einstein cylinder.
   AdS$_{4}$  is the region $0 \le \psi \le \pi/2$ and
   $-\pi \le t_{c} <  \pi$.
   The covering space-time, CAdS$_{4}$,
   is the region $-\infty < t_{c} < \infty$.
   The static Einstein universe is the region
   $0 \le \psi \le \pi$ and  $-\infty < t_{c} < \infty$.
   Time-like and null world-lines
   are indicated.
   The $45^{\circ}$ dashed lines which
   form the diamonds are the
   null geodesics. The dotted
   periodic lines are
   the time-like geodesics.
   For increasing energy,
   the time-like geodesics
   approach the nulls and thus reach closer to affine infinity
   located at
   $\psi = \pi/2$.  Left of the center is identified with
   $\theta = 0$ and the right with $\theta = \pi$.
   Data place on the constant time slices,
   say $t_{c} = -\pi/2$, have past and future
   Cauchy horizons given by the dashed
   nulls forming
   the diamonds.  In the horo-spherical coordinates
   with line element $ds^{2} =
   (\alpha z)^{-2}(dt^{2} - dx^{2} - dy^{2} - dz^{2})$,
   the affine boundary
   at $\psi = \pi/2$
   maps to $z=0$ [17,19,20]
   (Figure taken after Avis, Isham, and Storey [58]).}
\label{fig9}
\end{figure}

\end{document}